\documentclass[11pt]{article}
\usepackage{amsmath}   
\usepackage{amsfonts}    
\usepackage{mathrsfs} 
\usepackage{xfrac} 
\usepackage{tabu}
\usepackage{color}

\usepackage{tikz}
\usetikzlibrary{trees,snakes}
\usepackage{verbatim}

\usepackage{amsmath}
\usepackage{amsfonts}
\usepackage{amssymb}
\usepackage{graphicx}
\usepackage{tikz}
\usetikzlibrary{calc}
\usepackage{pgfplots}
\usepackage{cleveref}
\usepackage{graphicx}
\usepackage{subcaption}
\usetikzlibrary{trees,snakes}
\usepackage{verbatim}

\usepackage{mathrsfs} 
\usepackage{xfrac} 
\usepackage{tabu}
\usepackage{color}
\usepackage{tikz}
\usepackage{bbm}

\definecolor{myred}{RGB}{255, 0, 0}
\definecolor{myblue}{RGB}{0, 0, 255}
\newtheorem{theorem}{Theorem}
\newtheorem{lemma}{Lemma}
\newtheorem{proposition}{Proposition}

\newcommand {\nn} {\nonumber}
\newcommand {\pr} {\mathbb{P}}
\newcommand{\IND}{\mathbbm{1}}

\newcommand{\dfn}{\stackrel{\triangle}{=}}

\newcommand {\bx} {\boldsymbol{x}}
\newcommand {\by} {\boldsymbol{y}}

\newcommand {\bX} {\boldsymbol{X}}
\newcommand{\calA}{{\cal A}}
\newcommand{\calB}{{\cal B}}
\newcommand{\calC}{{\cal C}}

\newcommand{\calE}{{\cal E}}

\newcommand{\calM}{{\cal M}}
\newcommand{\calN}{{\cal N}}

\newcommand{\calS}{{\cal S}}

\newcommand{\calX}{{\cal X}}

\begin{document}
\thispagestyle{empty}
\title{Fast Convergence to Unanimity in Dense Erd\H{o}s--R\'enyi Graphs\footnote{
		This research has been funded in part by ETH Foundations of Data Science (ETH-FDS).}\\}
\author{\\Ran Tamir\\}
\maketitle
\begin{center}
Signal and Information Processing Laboratory \\
ETH Zurich, 8092 Zurich, Switzerland \\ 
Email: tamir@isi.ee.ethz.ch
\end{center}
\vspace{1.5\baselineskip}
\setlength{\baselineskip}{1.5\baselineskip}

\begin{abstract}
Majority dynamics on the binomial Erd\H{o}s--R\'enyi graph $\mathsf{G}(n,p)$ with $p=\lambda/\sqrt{n}$ is studied. In this process, each vertex has a state in $\{0,1\}$ and at each round, every vertex adopts the state of the majority of its neighbors, retaining its state in the case of a tie.	
It was conjectured by Benjamini {\it et al.} and proved by Fountoulakis {\it et al.} that this process reaches unanimity with high probability in at most four rounds. By adding some extra randomness and allowing the underlying graph to be drawn anew in each communication round, we improve on their result and prove that this process reaches consensus in only three communication rounds with probability approaching $1$ as $n$ grows to infinity. 
We also provide a converse result, showing that three rounds are not only sufficient, but also necessary. \\

\end{abstract}

\clearpage

\section{Introduction}

In this work, we continue a specific line of research on majority dynamics in dense random graphs \cite{Benjamini2016, Fountoulakis2020}. 
We analyze the performance of a simple majority-rule protocol solving a fundamental coordination problem in distributed computing - \emph{binary majority consensus}, in the presence of a probabilistic network model.  
In the binary consensus problem, to begin with, every agent is initially assigned some binary value, deterministically or probabilistically, referred to as the agent's initial opinion. 
The objective of a protocol that solves consensus is to have all agents eventually decide on the same opinion, thus achieving unanimity throughout the system. 
In binary majority consensus, if a majority of agents initially hold the same opinion, which is usually the case, then all agents must decide on this opinion. Majority consensus is utilized when, beyond facilitating agreement, the agreed upon opinion holds importance.
Consensus is an elementary problem in distributed computing, as many other coordination problems were shown to be directly reducible to and from consensus. The list includes agreeing on what transactions to commit to a database \cite{gray2006consensus}, state machine replication \cite{antoniadis2018state}, atomic snapshots \cite{attiya1998atomic}, total ordering of concurrent events \cite{lamport2019time}, and many more. 

We analyze the performance of the simple majority protocol (SMP) when the underlying network is governed by the Erd\H{o}s--R\'enyi random graph model $\mathsf{G}(n,p)$. In SMP, agents communicate in equal-length time intervals called communication rounds, or just rounds in short. All messages are sent at the beginning of a communication round, and they arrive by the end of the round.  
The SMP is shortly described as follows: In each round, every agent communicates its current opinion to all other agents that are connected to him through the underlying (random) graph. Then, it waits to receive all messages from other agents proposing their own opinions. If a majority of received messages advise the same opinion, then the agent adopts this opinion for the next round. All ties are solved by readopting the agent's own opinion. After a fixed, predefined number of communication rounds \textbf{\emph{r}}, each agent decides on its current opinion.

In \cite{Benjamini2016, Fountoulakis2020}, the binary majority consensus problem was solved for a $\mathsf{G}(n,p)$ random graph with a connectivity parameter $p \geq \tfrac{\lambda}{\sqrt{n}}$, for some sufficiently large $\lambda>0$ with random initial states, which is exactly the assumptions adopted in the current work. A remarkable result was conjectured in \cite{Benjamini2016} and proved in \cite{Fountoulakis2020}, stating that a majority consensus can be reached in at most four communication rounds with high probability. By allowing the underlying graph to be drawn independently for each round anew, we improve on this result, and prove that the SMP with $\boldsymbol{r}=3$ reaches majority consensus with probability converging to 1 as $n$ tends to infinity. 
We also show that this achievability result is tight. We will prove here that $\boldsymbol{r}=3$ communication rounds is a necessary condition, since the probability to reach unanimity with only $\boldsymbol{r}=2$ rounds converges to 0 as $n \to \infty$.    
We also study the exact dynamics of the system. 

\subsection{Related Work}
The problem of binary majority consensus was extensively studied in many different fields and contexts including autonomous systems \cite{mustafa2001majority, moreira2004efficient, gogolev2015distributed}, distributed systems \cite{thomas1979majority, breitwieser1982distributed, kanrar2016new}, and information theory \cite{mostofi2007binary, perron2009using, cruise2014probabilistic}, for a non-exhaustive list. 
Mustafa and Peke{\v{c}} \cite{mustafa2001majority}, studied the requirements on the connectivity of the network such that the SMP reaches unanimity for any initial assignment of agent opinions. The main result in \cite{mustafa2001majority} is that the SMP converges to the majority consensus successfully only in highly-connected networks. Our network model assumptions rely on their result.  
Our work closely resembles the work done in \cite{perron2009using, cruise2014probabilistic}. These papers have proved that in a fully-connected network where agents poll a portion of their neighbors uniformly at random, the SMP converges rapidly to majority consensus with probability of error (in the sense that unanimity was obtained, but not on the majority opinion) that tends to zero exponentially fast as $n \to \infty$. 

{Yet, another line of relatively recent work deserves a special attention.} 
In \cite{PAPER1}, a local polling protocol is proposed, and it is proved that it reaches consensus on the initial global majority in a random Erd\H{o}s--R\'enyi graph $\mathsf{G}(n,p)$ with $p=d/n$ where $d \geq (2+\epsilon)\log n$.    
An estimation on the number of required steps to reach consensus is provided. 
In \cite{PAPER2}, similar results were given for random regular graphs. In both of these papers, it is assumed that a clear bias exists between the two initial opinions.
In \cite{PAPER3}, the binary consensus problem was tackled from a different angle. For a random Erd\H{o}s--R\'enyi graph $\mathsf{G}(n,p)$ with a connectivity parameter $p \in (0,1)$ and any given $\epsilon \in (0,1)$, this work reveals what the initial difference between the two camps should be, such that the larger camp will eventually win with probability at least as high as $1-\epsilon$.       
Also for a random graph $\mathsf{G}(n,p)$ with $p=d/n$, it was proved in \cite{PAPER4} that if the probability assignment on one of the initial states behaves like $\tfrac{1}{2}+\omega\left(\tfrac{1}{\sqrt{d}}\right)$ and $d>(1+\epsilon)\log n$, then with high probability the process reaches unanimity in a constant number of rounds.

In \cite{TLS}, the performance of the SMP is analyzed in the presence of a probabilistic message loss. 
It is proved that in a fully-connected network, the SMP reaches consensus in only three communication rounds with probability converging to $1$ as $n \to \infty$, regardless of the initial state. 
It is proved in \cite{TLS} that if the difference between the numbers of agents that initially hold different opinions grows at a rate of $\sqrt{n}$, then the SMP with two communication rounds reaches unanimity on the majority opinion of the network, and if this difference grows faster than $\sqrt{n}$, then the SMP attains consensus on the majority opinion of the network in a single round, with probability approaching $1$ exponentially fast as $n \rightarrow \infty$.  

The remaining part of the paper is organized as follows. 
In Section \ref{sec2}, we establish notation conventions. 
In Section \ref{sec3}, we formalize the model, the protocol, and the objectives of this work. 
In Section \ref{sec4}, we provide and discuss the main results of this work, and in Sections \ref{sec5} and \ref{sec6}, we prove them.

\section{Notation Conventions} \label{sec2}

Throughout the paper, random variables will be denoted by capital letters, realizations will be denoted by the corresponding lower case letters, and their alphabets will be denoted by calligraphic letters. 
Random vectors and their realizations will be denoted, 
respectively, by boldface capital and lower case letters. 
Their alphabets will be superscripted by their dimensions. 
The binary Kullback-Leibler divergence function between two binary probability distributions with parameters $\alpha,\beta \in [0,1]$ is defined as
\begin{align} \label{DEF_Bin_DIVERGENCE}
D(\alpha \| \beta) = \alpha \log \left(\frac{\alpha}{\beta}\right) + (1-\alpha) \log \left(\frac{1-\alpha}{1-\beta}\right),
\end{align}
where logarithms, here and throughout the sequel, are understood to be taken to the natural base.
The probability of an event $\calE$ will be denoted by $\pr\{\calE\}$, and the expectation operator by $\mathbb{E}[\cdot]$.
The variance of a random variable $X$ is denoted by $\text{Var}[X]$. 
The indicator function of an event $\calA$ 
will be denoted by $\IND\{\calA\}$. 
The notation $[x]_{+}$ will stand for $\max\{0,x\}$.
For $\bx = (x_{1},x_{2},\ldots,x_{n}) \in \calX^{n}$ and for any $a \in \calX$, let us denote
\begin{align}
N(\bx;a) = \sum_{i=1}^{n} \IND\{x_{i}=a\}.
\end{align}

Let us denote by $\text{Ber}(p)$ a Bernoulli random variable with a success probability $p$ and by $\text{Bin}(n,p)$ a binomial random variable with $n$ independent experiments, each one with a success probability $p$. We adopt the following convention: if an event contains at least 2 binomial random variables, then we assume that they are statistically independent.

We will be concerned with the Erd\H{o}s--R\'enyi random graph model, to be denoted by $\mathsf{G}(n,p)$. In this model, a graph over $n$ vertices is constructed by connecting vertices randomly. Each edge is included in the graph with probability $p$ independent from every other edge.

\section{Model, Protocol, and Objectives} \label{sec3}

Assume a set of $2n$ agents, and denote their assignment of initial opinions by $\bx_{0,n} \in \{0,1\}^{2n}$. The vector $\bx_{0,n}$ is called the initial state. 
We assume that each agent picks his initial opinion at random according to a $\text{Ber}(\tfrac{1}{2})$ random variable. All the initial opinions are independent. 
Denote the numbers of zeros and ones in $\bx_{0,n}$ by $\mathsf{I}_{0}$ and $\mathsf{I}_{1}$, respectively. 
At each round, the $2n$ agents are randomly connected according to a $\mathsf{G}(2n,\tfrac{\lambda}{\sqrt{n}})$ graph, where $\lambda>0$ is independent of $n$. 
In every round, each agent transmits its current state over the connected network.

At round $\ell \geq 1$, assume that agent $i$ receives opinions from other $c(\ell,i)$ different agents, whose indexes are given by $(a_{1},a_{2},\ldots,a_{c(\ell,i)})$. 
The agent $i \in \{1,2,\ldots,2n\}$ receives the (random) vector: 
\begin{align}
\by_{\ell}^{i} = (y_{\ell}^{i}(a_{1}),y_{\ell}^{i}(a_{2}), \ldots, y_{\ell}^{i}(a_{c(\ell,i)})) \in \{0,1\}^{c(\ell,i)},
\end{align}
and for $b \in \{0,1\}$, he calculates the enumerators:
\begin{align}
\mathsf{N}_{\ell,i}(b) = \sum_{j=1}^{c(\ell,i)} \IND\{y_{\ell}^{i}(a_{j}) = b\}.
\end{align}
In the simple majority protocol (SMP), each agent updates its value according to the more common value at hand, i.e., agent $i$ chooses:
\begin{align}
\bx_{\ell}(i)
= \left\{   
\begin{array}{l l}
0    & \quad \text{if        $\mathsf{N}_{\ell,i}(0) > \mathsf{N}_{\ell,i}(1)$}   \\
1    & \quad \text{if        $\mathsf{N}_{\ell,i}(0) < \mathsf{N}_{\ell,i}(1)$}   \\
\bx_{\ell-1}(i)    & \quad \text{if        $\mathsf{N}_{\ell,i}(0) = \mathsf{N}_{\ell,i}(1)$}
\end{array} \right. .
\end{align} 
The vector $\bx_{\ell} \in \{0,1\}^{2n}$ is called the state at the end of round $\ell$.   

A specific SMP defines a-priori the number of rounds until termination. 
Let us denote by SMP$(r)$ the SMP with $r$ rounds of communication until termination.
We say that the SMP$(r)$ attains {\it consensus} if 
\begin{align}
\bx_{r}(1) = \bx_{r}(2) = \ldots = \bx_{r}(2n), 
\end{align}  
and denote this event by $\mathsf{Con}(r,n)$. 
Similarly, we say that the SMP$(r)$ attains {\it majority consensus} if the following holds: 
\begin{align}
\mathsf{I}_{0} > \mathsf{I}_{1} ~~ &\rightarrow ~~ 
\bx_{r}(1) = \bx_{r}(2) = \ldots = \bx_{r}(2n) = 0, \\
\mathsf{I}_{0} < \mathsf{I}_{1} ~~ &\rightarrow ~~ 
\bx_{r}(1) = \bx_{r}(2) = \ldots = \bx_{r}(2n) = 1, \\
\mathsf{I}_{0} = \mathsf{I}_{1} ~~ &\rightarrow ~~ 
\bx_{r}(1) = \bx_{r}(2) = \ldots = \bx_{r}(2n),
\end{align}  
and denote this event by $\mathsf{MCon}(r,n)$.

Now, the objective of this work is to prove that the SMP requires only three rounds of communication in order to attain consensus, with a probability that converges to 1 when $n \to \infty$.

\section{Main Results} \label{sec4}

Our first main result in this work is as follows.

\begin{theorem} \label{Main_THEOREM}
	Let $2n$ agents draw their initial opinions using independent $\mathsf{Ber}(\tfrac{1}{2})$ random variables. Assume that the $2n$ agents communicate over a graph $\mathsf{G}(2n,\tfrac{\lambda}{\sqrt{n}})$, with $\lambda>0$, which is drawn independently at each round. Then, it holds that $\pr\{\mathsf{MCon}(3,n)\} \to 1$ as $n \to \infty$. 
\end{theorem}

\subsection*{Discussion}

Beyond the simple fact that consensus is attained in only three communication rounds, with a probability that converges to one as $n \to \infty$, we also study the exact dynamics of the system in each of the intermediate steps. At the beginning, each agent tosses a fair coin to determine her initial opinion. We prove in Proposition \ref{PROP_1} that with high probability, one of the two opinions will have a majority of the order of $\sqrt{n}$, e.g., about $n+\alpha\sqrt{n}$ agents will hold the opinion 0 and $n-\alpha\sqrt{n}$ agents will hold the opinion 1, for some $\alpha>0$. Then, under the assumption that the initial state have a majority of zeros, we prove in Proposition \ref{PROP_3} that with high probability, at the end of the first communication round, the agents holding the opinion zero will have a bigger majority, which is of the order of $n^{3/4}$. In other words, about $n+\beta n^{3/4}$ agents will hold the opinion 0 and $n-\beta n^{3/4}$ agents will hold the opinion 1, for some $\beta>0$. Moving further, we prove in Proposition \ref{PROP_5} that after the second communication round, with a probability converging to 1 exponentially fast, the agents holding the opinion zero will have a linear majority, i.e., about $n+\gamma n$ agents will hold the opinion 0 and $n-\gamma n$ agents will hold the opinion 1, for some $\gamma>0$. Then, only one more communication round is required to reach consensus; we prove in Proposition \ref{PROP_6} that after the third communication round, again with a probability converging to 1 exponentially fast, all agents will hold the opinion zero (or one, if the initial majority was in favor of the ones). For clarity, the dynamics of the system is demonstrated in Figure \ref{fig:Evolution} below. 

It was proved by Benjamini {\it et al.} \cite[Theorem 2]{Benjamini2016} that majority consensus is attained after four communication rounds with probability slightly higher than 0.4 over the choice of the initial states and over the choice of the random graph. It was assumed in \cite{Benjamini2016} that the underlying graph is chosen only once and remains fixed. It was conjectured in \cite{Benjamini2016} that, in fact, majority consensus can be reached with high probability as $n \to \infty$, and this conjecture was recently proved by Fountoulakis {\it et al.} \cite{Fountoulakis2020}. The proof in \cite{Fountoulakis2020} consists of two main parts; the first part involves with the first two rounds and the second part with the last two rounds. It is assumed that the underlying random graph is drawn twice - before the first round and between the second and the third rounds. Here, in Theorem \ref{Main_THEOREM}, we allow the underlying network to be drawn before each communication round starts, and this enables to reach consensus faster than in \cite{Benjamini2016, Fountoulakis2020}; the entire community agrees on the initial majority after only three rounds with probability converging to 1 as $n \to \infty$. It is important to note that Theorem \ref{Main_THEOREM} holds for any constant $\lambda>0$, while the results in \cite{Benjamini2016, Fountoulakis2020} hold with $\lambda$ being a sufficiently large universal constant.    

In Theorem \ref{Main_THEOREM} above, we assume that the underlying graph is drawn independently for each communication round as $\mathsf{G}(2n,\tfrac{\lambda}{\sqrt{n}})$, with $\lambda>0$. This means that each agent communicate her current state to around $\lambda\sqrt{n}$ other agents. It may be questioned whether consensus can be reached when communicating over sparser networks. As long as consensus is attained in three rounds, only a partial relaxation can be made. By examining the proof of Theorem \ref{Main_THEOREM}, it seems that if the graph in the third round is $\mathsf{G}(2n,\tfrac{\lambda}{n^{\xi}})$, $\xi \in (\tfrac{1}{2},1)$, without altering the random graphs at the first two rounds, then consensus can still be reached in three rounds with probability converging to 1 as $n \to \infty$. We conjecture that a majority consensus can be attained when the underlying graph at all communication rounds is $\mathsf{G}(2n,\tfrac{\lambda}{n^{\xi}})$, $\xi \in (\tfrac{1}{2},1)$, but then, the number of required rounds should be higher than three. We leave this point open for future research.

The result provided in Theorem \ref{Main_THEOREM} is, in fact, an achievability result, i.e., it only tells under what conditions consensus can be attained. Hence, it is worth investigating whether consensus may be attained by the SMP with even less communication rounds than required in Theorem \ref{Main_THEOREM}.     
In the following result, which is the second main result of this work and is proved in Section \ref{sec6}, we show that three rounds of communications are not only sufficient, but also necessary.

\begin{theorem} \label{Main_THEOREM2}
	Let $2n$ agents draw their initial opinions using independent $\mathsf{Ber}(\tfrac{1}{2})$ random variables. Assume that the $2n$ agents communicate over a graph $\mathsf{G}(2n,\tfrac{\lambda}{\sqrt{n}})$, with $\lambda>0$, which is drawn independently at each round. Then, it holds that $\pr\{\mathsf{Con}(2,n)\} \to 0$ as $n \to \infty$. 
\end{theorem}

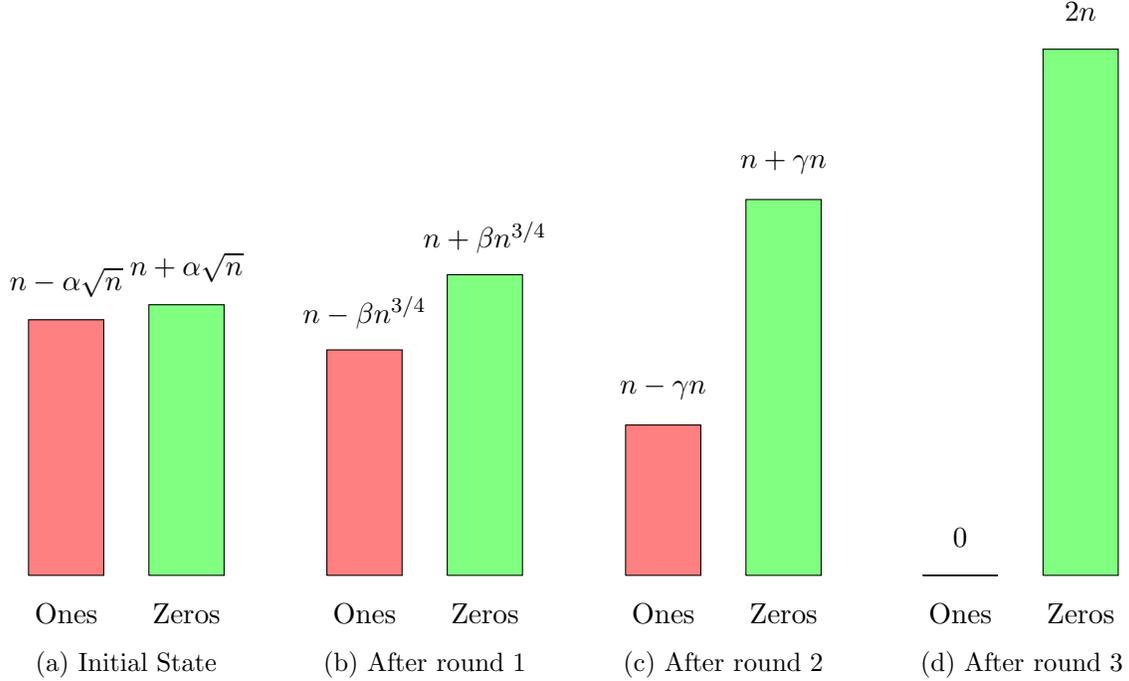
\begin{figure}[h!]
	\definecolor{CODEWORD}{rgb}{0,0,0}
	\definecolor{FULL}{rgb}{0.13,0.65,0.13}
	\definecolor{EMPTY}{rgb}{1.0,0,0.22}	
	\begin{subfigure}[b]{0.25\columnwidth}
		\centering 
		\begin{tikzpicture}
		\draw[fill=red!50!white] (0,0) -- (0,3.4) -- (1,3.4) -- (1,0) -- (0,0);
		\draw[fill=green!50!white] (1.6,0) -- (1.6,3.6) -- (2.6,3.6) -- (2.6,0) -- (1.6,0);
		\node at (0.5,3.9) {$n-\alpha\sqrt{n}$};
		\node at (2.1,4.1) {$n+\alpha\sqrt{n}$};
		\node at (0.5,-0.5) {Ones};
		\node at (2.1,-0.5) {Zeros};
		\end{tikzpicture}
		\caption{Initial State} 
		\label{fig:POPULATION1}
	\end{subfigure}%
	\begin{subfigure}[b]{0.25\columnwidth}
		\centering 
		\begin{tikzpicture}
		\draw[fill=red!50!white] (0,0) -- (0,3) -- (1,3) -- (1,0) -- (0,0);
		\draw[fill=green!50!white] (1.6,0) -- (1.6,4) -- (2.6,4) -- (2.6,0) -- (1.6,0);
		\node at (0.5,3.5) {$n-\beta n^{3/4}$};
		\node at (2.1,4.5) {$n+\beta n^{3/4}$};
		\node at (0.5,-0.5) {Ones};
		\node at (2.1,-0.5) {Zeros};
		\end{tikzpicture}
		\caption{After round 1} 
		\label{fig:POPULATION2}
	\end{subfigure}%
	\begin{subfigure}[b]{0.25\columnwidth}
		\centering 
		\begin{tikzpicture}
		\draw[fill=red!50!white] (0,0) -- (0,2) -- (1,2) -- (1,0) -- (0,0);
		\draw[fill=green!50!white] (1.6,0) -- (1.6,5) -- (2.6,5) -- (2.6,0) -- (1.6,0);
		\node at (0.5,2.5) {$n-\gamma n$};
		\node at (2.1,5.5) {$n+\gamma n$};
		\node at (0.5,-0.5) {Ones};
		\node at (2.1,-0.5) {Zeros};
		\end{tikzpicture}
		\caption{After round 2} 
		\label{fig:POPULATION3}
	\end{subfigure}%
	\begin{subfigure}[b]{0.25\columnwidth}
		\centering 
		\begin{tikzpicture}
		\draw[fill=red!50!white] (0,0) -- (0,0) -- (1,0) -- (1,0) -- (0,0);
		\draw[fill=green!50!white] (1.6,0) -- (1.6,7) -- (2.6,7) -- (2.6,0) -- (1.6,0);
		\node at (0.5,0.5) {$0$};
		\node at (2.1,7.5) {$2n$};
		\node at (0.5,-0.5) {Ones};
		\node at (2.1,-0.5) {Zeros};
		\end{tikzpicture}
		\caption{After round 3} 
		\label{fig:POPULATION4}
	\end{subfigure}%
	\caption{Typical evolution in 3 communication rounds.} 
	\label{fig:Evolution}
\end{figure}

\section{Proof of Theorem \ref{Main_THEOREM}} \label{sec5}
\subsection{At the Initial State}
The following proposition shows that if every agent picks his initial opinion at random with a fair coin, then the initial state of the entire community will be asymmetric of order at least $\sqrt{n}$. This result is proved in Appendix A.

\begin{proposition} \label{PROP_1}
	Let $\bx_{0,n} \in \{0,1\}^{2n}$ be a sequence of random initial states, where each agent tosses a fair coin to determine his initial opinion. Let $\epsilon > 0$ be given. Then, there exist $\alpha=\alpha(\epsilon)$ with $\alpha(\epsilon) \xrightarrow{\epsilon \to 0} 0$ and $M(\epsilon)$, such that for all $n \geq M(\epsilon)$,
	\begin{align}
	\pr \left\{ \{N(\bX_{0};0) \leq n - \alpha\sqrt{n}\}
	\cup \{N(\bX_{0};0) \geq n + \alpha\sqrt{n}\} \right\} \geq 1 - \epsilon.
	\end{align} 
\end{proposition}

\subsection{At the First Round}

In the following result, which is proved in Appendix B, we show that if the difference between the number of agents holding `zero' and the number of agents holding `one' is at the order of $\sqrt{n}$, then the probability that any agent will update its state to `zero' converges to $\tfrac{1}{2}$ as $n \to \infty$, but at a relatively slow rate. This fact is important for increasing the difference between the two opinions after the first round of communication.    

\begin{proposition} \label{PROP_2}
	Let $\bx_{0,n} \in \{0,1\}^{2n}$ be a sequence of initial states with $n+\alpha\sqrt{n}$ zeros and $n-\alpha\sqrt{n}$ ones and assume that the underlying graph is $\mathsf{G}(2n,\tfrac{\lambda}{\sqrt{n}})$. 
	Define the constant 
	\begin{align}
	C_{0}(\alpha,\lambda)=\frac{\pi}{8e^{4}}
	\frac{([\alpha\lambda-\sqrt{2\alpha\lambda}]_{+}+1)\exp\{-4/\lambda\}}{\lambda}.
	\end{align}
	If an agent starts with a `0' or a `1', then the probability to update to `0' is lower-bounded as 
	\begin{align} 
	Q_{n} \geq \frac{1}{2} + \frac{C_{0}(\alpha,\lambda)}{n^{1/4}}.
	\end{align} 
\end{proposition}

Based on the result of Proposition \ref{PROP_2}, we next state that if the difference between the two different opinions at the initial state is at the order of $\sqrt{n}$, then with probability converging to 1 as $n \to \infty$, after one round of communication, the difference will grow to the order of $n^{3/4}$. The following result, which is based on Chernoff's inequality, is proved in Appendix D.

\begin{proposition} \label{PROP_3}
	Let $\bx_{0,n} \in \{0,1\}^{2n}$ be a sequence of initial states with $n+\alpha\sqrt{n}$ zeros and $n-\alpha\sqrt{n}$ ones and assume that the underlying graph is $\mathsf{G}(2n,\tfrac{\lambda}{\sqrt{n}})$. 
	Then,
	\begin{align}
	\pr \left\{N(\bX_{1};0) \geq  n + C_{0}(\alpha,\lambda) n^{3/4} \right\}
	&\geq 1 - \exp\left\{- C_{0}(\alpha,\lambda)^{2} \sqrt{n} \right\} \xrightarrow{n \to \infty} 1.
	\end{align} 
\end{proposition}

\subsection{At the Second Round}

In the following result, which is proved in Appendix E, we show that if the difference between the number of agents holding opinion `zero' and the number of agents holding opinion `one' is at the order of $n^{3/4}$, then the probability that an agent will update its state to `zero' is bounded away from $\tfrac{1}{2}$ for any $n$. Due to this fact, the difference between the two opinions after the second round of communication is going to be linear in $n$.

\begin{proposition} \label{PROP_4}
	Let $\bx_{0,n} \in \{0,1\}^{2n}$ be a sequence of initial states with $n+\beta n^{\frac{3}{4}}$ zeros and $n-\beta n^{\frac{3}{4}}$ ones and assume that the underlying graph is $\mathsf{G}(2n,\tfrac{\lambda}{\sqrt{n}})$.
	Define the constant 
	\begin{align}
	C_{1}(\beta,\lambda) = \frac{(\pi\beta)^{\frac{3}{2}} \exp\{-(4+2\beta)/\lambda\} \exp\left\{-4\beta^{2}\lambda\right\}}{e^{6}\sqrt{\lambda}}. 
	\end{align}
	If an agent starts with a `0' or a `1', then the probability to update to `0' is lower-bounded as 
	\begin{align} \label{Prop4_Res}
	Q_{n} \geq \frac{1}{2} + C_{1}(\beta,\lambda).
	\end{align} 
\end{proposition} 

Based on the result of Proposition \ref{PROP_4}, we next state that if the difference between the two different opinions is at the order of $n^{3/4}$, then with probability converging to 1 exponentially fast as $n \to \infty$, after one round of communication, the difference will grow to the order of $n$.

\begin{proposition} \label{PROP_5}
	Let $\bx_{0,n} \in \{0,1\}^{2n}$ be a sequence of initial states with $n+\beta n^{\frac{3}{4}}$ zeros and $n-\beta n^{\frac{3}{4}}$ ones and assume that the underlying graph is $\mathsf{G}(2n,\tfrac{\lambda}{\sqrt{n}})$. 
	Then,
	\begin{align}
	\pr \left\{N(\bX_{1};0) \geq n + C_{1}(\beta,\lambda) n \right\} 
	\geq 1 - \exp\left\{-C_{1}(\beta,\lambda)^{2} n \right\}
	\xrightarrow{n \to \infty} 1.
	\end{align} 
\end{proposition} 
We omit the proof of Proposition \ref{PROP_5}, since it is very similar to the proof of Proposition \ref{PROP_3}.

\subsection{At the Third Round}

According to Proposition \ref{PROP_5}, after the second round of communication, the number of agents holding opinion `zero' (or `one') is going to be larger by $\gamma n$ than the number of agents holding opinion `one', with a probability converging to 1 as $n \to \infty$. Then, in the following result, which is proved in Appendix F, we state that only one more round is required in order to reach consensus.

\begin{proposition} \label{PROP_6}
	Let $\bx_{0,n} \in \{0,1\}^{2n}$ be a sequence of initial states with $n+\gamma n$ zeros and $n-\gamma n$ ones, where $\gamma \in (0,1)$, and assume that the underlying graph is $\mathsf{G}(2n,\tfrac{\lambda}{\sqrt{n}})$.
	Then, the SMP$(1)$ reaches consensus with probability converging to 1 as $n \to \infty$. Specifically,  
	\begin{align}
	\label{prop6_res}
	\pr\left\{N(\bX_{1};0) = 2n \right\}
	&\geq 1 - 2n \sqrt{\tfrac{1+\gamma}{1-\gamma}} \cdot \exp \left\{-\lambda \gamma^{2} \sqrt{n} \right\}.
	\end{align}
\end{proposition}

The fact that the probability in \eqref{prop6_res} converges to 1 relatively fast suggest that consensus may be attained even if the network at the third communication round is sparser. Indeed, by examining the proof of Proposition \ref{PROP_6}, we find out that if the underlying graph is $\mathsf{G}(2n,\tfrac{\lambda}{n^{\xi}})$, $\xi \in (\tfrac{1}{2},1)$, then \eqref{prop6_res} is generalized to   
\begin{align}
\label{prop6_res_b}
\pr\left\{N(\bX_{1};0) = 2n \right\}
&\geq 1 - 2n \sqrt{\tfrac{1+\gamma}{1-\gamma}} \cdot \exp \left\{-\lambda \gamma^{2} n^{1-\xi} \right\},
\end{align}
which still converges to 1 for any $\xi \in (\tfrac{1}{2},1)$. It is important to note that it is impossible to reach consensus with high probability in three communication rounds when the underlying graph at all rounds is $\mathsf{G}(2n,\tfrac{\lambda}{n^{\xi}})$, $\xi \in (\tfrac{1}{2},1)$. Under such conditions, the required number of rounds for attaining consensus is greater than 3, and this is left open for future research.

\subsection{Wrapping Up}

We are now able to prove Theorem \ref{Main_THEOREM}. 
Let $\epsilon_{1},\epsilon_{2},\epsilon_{3},\epsilon_{4} > 0$ be arbitrarily small given constants. 

Let $\alpha$ be as in Proposition \ref{PROP_1} corresponding to $\epsilon_{1}$ and define the following event
\begin{align}
\calA_{n} = \left\{N(\bX_{0};0) \leq n - \alpha \sqrt{n} \text{~~or~~} N(\bX_{0};0) \geq n + \alpha \sqrt{n} \right\},
\end{align}
such that $\pr\{\calA_{n}\} \geq 1-\epsilon_{1}$ for all sufficiently large $n$.

Next, for a given $\alpha$ and $\lambda$, choose $\beta=C_{0}(\alpha,\lambda)$ as defined in Proposition \ref{PROP_2}, and define the event
\begin{align}
\calB_{n} = \left\{N(\bX_{1};0) \leq n-\beta n^{\frac{3}{4}} \text{~~or~~} N(\bX_{1};0) \geq n+\beta n^{\frac{3}{4}} \right\},
\end{align} 
such that $\pr\{\calB_{n}|\calA_{n}\} \geq 1-\epsilon_{2}$ for all sufficiently large $n$, according to Proposition \ref{PROP_3}.

Furthermore, for a given $\beta$ and $\lambda$, choose $\gamma=C_{1}(\beta,\lambda)$ as defined in Proposition \ref{PROP_4}, and define the event
\begin{align}
\calC_{n} = \left\{N(\bX_{2};0) \leq n-\gamma n \text{~~or~~} N(\bX_{2};0) \geq n+\gamma n \right\},
\end{align}
such that $\pr\{\calC_{n}|\calB_{n}\} \geq 1-\epsilon_{3}$ for all sufficiently large $n$, according to Proposition \ref{PROP_5}.

Then, consider the following.
\begin{align}
\pr\{\mathsf{Con}(3,n)\} 
&= \pr\{ N(\bX_{3};0) = 0 \text{~or~} N(\bX_{3};0) = 2n\}\\
\label{A_To_exp1}
&= \pr\{ N(\bX_{3};0) = 0 \text{~or~} N(\bX_{3};0) = 2n| \calC_{n}\} \cdot \pr\{\calC_{n}\} \nn \\
&~~+ \pr\{ N(\bX_{3};0) = 0 \text{~or~} N(\bX_{3};0) = 2n| \calC_{n}^{\mbox{\tiny c}}\} \cdot \pr\{\calC_{n}^{\mbox{\tiny c}}\}  \\
&\geq \pr\{ N(\bX_{3};0) = 0 \text{~or~} N(\bX_{3};0) = 2n| \calC_{n}\} \cdot \pr\{\calC_{n}\}  \\
\label{A_To_exp2}
&\geq \left(1 - \epsilon_{4}\right) \cdot \pr\{\calC_{n}\},
\end{align}
where \eqref{A_To_exp1} follows from the law of total probability and \eqref{A_To_exp2} holds for all large enough $n$, due to Proposition \ref{PROP_6}. Furthermore,
\begin{align}
\pr\{\calC_{n}\} 
\label{A_To_exp3}
&= \pr\{ \calC_{n}| \calB_{n}\} \cdot \pr\{\calB_{n}\} + \pr\{ \calC_{n}| \calB_{n}^{\mbox{\tiny c}}\} \cdot \pr\{\calB_{n}^{\mbox{\tiny c}}\}  \\
&\geq \pr\{ \calC_{n}| \calB_{n}\} \cdot \pr\{\calB_{n}\} \\
\label{A_To_exp4}
&\geq (1-\epsilon_{3}) \cdot \pr\{\calB_{n}\} \\
\label{A_To_exp5}
&\geq (1-\epsilon_{3}) \cdot \pr\{\calB_{n}|\calA_{n}\} \cdot \pr\{\calA_{n}\}\\
\label{A_To_exp6}
&\geq (1-\epsilon_{3}) \cdot (1-\epsilon_{2}) \cdot (1-\epsilon_{1}),
\end{align}
where \eqref{A_To_exp3} is due to the law of total probability, \eqref{A_To_exp4} follows from Proposition \ref{PROP_5} for all $n$ sufficiently large, and \eqref{A_To_exp5} follows from the law of total probability.  
The passage to \eqref{A_To_exp6} follows from Propositions \ref{PROP_1} and \ref{PROP_3}, for all $n$ sufficiently large. 
Substituting \eqref{A_To_exp6} back into \eqref{A_To_exp2}, we conclude that $\pr\{\mathsf{Con}(3,n)\}$ can be made arbitrarily close to 1, which proves Theorem \ref{Main_THEOREM}.

\section{Proof of Theorem \ref{Main_THEOREM2}}  \label{sec6}

\subsection{Main Ingredients}

In the following negative result, which is proved in Appendix G, we show that even if the number of agents holding opinion `zero' is greater than the number of agents holding opinion `one', then the number of `zero' agents cannot grow too much in a single communication round.  

\begin{proposition} \label{PROP_7}
	Let $\{A_{n}\}_{n=1}^{\infty}$ and $\{B_{n}\}_{n=1}^{\infty}$ be two monotonically increasing sequences with $B_{n} \geq A_{n}$ for every $n$.
	Let $\bx_{0,n} \in \{0,1\}^{2n}$ be a sequence of initial states with $n+A_{n}$ zeros and $n-A_{n}$ ones and assume that if an agent starts with a `0' or a `1', then the probability to update to `0' is upper-bounded by $P_{n}$. Then, it holds that
	\begin{align}
	\pr \left\{N(\bX_{1};0) \geq n + B_{n}\right\} \leq \exp \left\{-2n \cdot D\left(\frac{1}{2} + \frac{B_{n}}{2n} \middle\| P_{n} \right) \right\}.
	\end{align} 
\end{proposition} 

While in Proposition \ref{PROP_2} we stated a positive result that if the difference between the two camps is at the order of $\sqrt{n}$, then the probability of updating to `zero' is {\it lower-bounded} by $\tfrac{1}{2}+\tfrac{C_{0}}{n^{1/4}}$, in the following result, which is proved in Appendix H, we state a negative result, which provides an upper bound on this probability. In the following result, we allow the difference between the two opinion numbers to be a general function of $n$.

\begin{proposition} \label{PROP_8}
	Let $\{\psi_{n}\}_{n=1}^{\infty}$ be a sequence such that $\lim_{n \to \infty} \frac{\psi_{n}}{n} = 0$.
	Let $\bx_{0,n} \in \{0,1\}^{2n}$ be a sequence of initial states with $n+\psi_{n}$ zeros and $n-\psi_{n}$ ones and assume that the underlying graph is $\mathsf{G}(2n,p_{n})$, $p_{n}=\tfrac{\lambda}{\sqrt{n}}$. Assume that $\psi_{n}p_{n}\geq 1$ for every $n$. 
	Define the sequence $\{\delta_{n}\}_{n=1}^{\infty}$ according to 
	\begin{align}
	n^{\delta_{n}} = \sqrt{\log(n^{\theta})},~~\theta > 5.
	\end{align}
	If an agent starts with a `0' or a `1', then for all sufficiently large $n$, the probability to update to `0' is upper-bounded as 
	\begin{align} 
	P_{n} \leq \frac{1}{2} + \frac{60\psi_{n}p_{n}}{\lambda n^{1/4-\delta_{n}}}.
	\end{align} 
\end{proposition} 

Proposition \ref{PROP_8} shows that if the difference between the two camps is again at the order of $\sqrt{n}$, then the probability of updating to `zero' is now {\it upper-bounded} by $\tfrac{1}{2}+\tfrac{k_{n}}{n^{1/4}}$, where $k_{n}$ grows only logarithmically fast in $n$. 
In the sequel, the result of Proposition \ref{PROP_8} will be instrumental in Proposition \ref{PROP_7}, which requires an upper bound on the probability of updating to `zero'. 

In Proposition \ref{PROP_6} we stated a positive result, according to which, if the difference between the number of opinions is at the order of $n$, then the probability to reach consensus in a single round converges to 1 as $n \to \infty$.  
In the following result, which is proved in Appendix J, we state a negative result on attaining consensus in a single communication round; if the difference between the two camps is too small, then reaching consensus is impossible in a single round.  

\begin{proposition} \label{PROP_9}
	Let $\{C_{n}\}_{n=1}^{\infty}$ be a sequence such that $\lim_{n \to \infty} \frac{C_{n}}{n} = 0$. Let $\bx_{0,n} \in \{0,1\}^{2n}$ be a sequence of initial states with $n+C_{n}$ zeros or $n+C_{n}$ ones, and assume that the underlying graph is $\mathsf{G}(2n,\tfrac{\lambda}{\sqrt{n}})$. 
	Then, the SMP$(1)$ is characterized by
	\begin{align}
	\label{prop9_res}
	\pr\{\mathsf{Con}(1,n)\} 
	\leq \exp\left\{ -\frac{n C_{n}^{2}}{2(n+C_{n})}
	\exp \left\{ - 32\lambda \cdot  \frac{C_{n}^{2}}{\sqrt{n}(n-C_{n})} \right\} \right\}.
	\end{align}
\end{proposition}

Specifically, if the difference between the number of opinions grows faster than $n^{3/4}$, then the inner exponent on the right-hand-side of \eqref{prop9_res} is relatively close to 0 and the bound is close to 1, which is useless. On the other hand, if the difference between the two camps grows slower than $n^{3/4}$, then the probability of reaching consensus converges to 0 as $n \to \infty$.

\subsection{Wrapping Up}

We are now in a good position to prove Theorem \ref{Main_THEOREM2}.
\begin{align}
\pr\{\mathsf{Con}(2,n)\} 
&= \pr\{ N(\bX_{2};0) = 0 \text{~or~} N(\bX_{2};0) = 2n\}\\
\label{ref80}
&\leq \pr\{N(\bX_{2};0) = 0\} + \pr\{N(\bX_{2};0) = 2n\}.
\end{align}
We will prove that the second term in \eqref{ref80} converges to zero as $n \to \infty$, while a proof for the first term follows similar lines. Consider the following:  
\begin{align}
\pr\{N(\bX_{2};0) = 2n\}
&= \sum_{\ell=0}^{2n} \pr\left\{N(\bX_{2};0) = 2n,N(\bX_{1};0) = \ell\right\} \\
&= \sum_{\ell=0}^{n+\sigma_{n}} \pr\left\{N(\bX_{2};0) = 2n,N(\bX_{1};0) = \ell\right\} \nn \\
&~~~~~~+\sum_{\ell=n+\sigma_{n}}^{2n} \pr\left\{N(\bX_{2};0) = 2n,N(\bX_{1};0) = \ell\right\} \\
&\leq \sum_{\ell=0}^{n+\sigma_{n}} \pr\left\{N(\bX_{2};0) = 2n,N(\bX_{1};0) = n+\sigma_{n} \right\} \nn \\
&~~~~~~+\sum_{\ell=n+\sigma_{n}}^{2n} \pr\left\{N(\bX_{1};0) = \ell\right\} \\
&= (n+\sigma_{n}+1) \cdot \pr\left\{N(\bX_{2};0) = 2n,N(\bX_{1};0) = n+\sigma_{n} \right\} \nn \\
&~~~~~~+\pr\left\{N(\bX_{1};0) \geq n+\sigma_{n}\right\}.
\end{align}
Since 
\begin{align}
&\pr\left\{N(\bX_{2};0) = 2n,N(\bX_{1};0) = n+\sigma_{n} \right\} \nn \\
&~~~= \pr\left\{N(\bX_{2};0) = 2n|N(\bX_{1};0) = n+\sigma_{n} \right\} \cdot \pr\left\{N(\bX_{1};0) = n+\sigma_{n} \right\} \\
&~~~\leq \pr\left\{N(\bX_{2};0) = 2n|N(\bX_{1};0) = n+\sigma_{n} \right\},
\end{align}
we arrive at
\begin{align}
\pr\{N(\bX_{2};0) = 2n\}
\label{ref81}
&\leq (n+\sigma_{n}+1) \cdot \pr\left\{N(\bX_{2};0) = 2n|N(\bX_{1};0) = n+\sigma_{n} \right\} \nn \\
&~~~~~~~~+\pr\left\{N(\bX_{1};0) \geq n+\sigma_{n}\right\}.
\end{align}
As for the second term in \eqref{ref81}, consider the following:
\begin{align}
\pr\{N(\bX_{1};0) \geq n+\sigma_{n}\}
&= \sum_{k=0}^{2n} \pr\left\{N(\bX_{1};0) \geq n+\sigma_{n},N(\bX_{0};0) = k\right\} \\
&= \sum_{k=0}^{n+\tau_{n}} \pr\left\{N(\bX_{1};0) \geq n+\sigma_{n},N(\bX_{0};0) = k\right\} \nn \\
&~~~~~~+\sum_{k=n+\tau_{n}}^{2n} \pr\left\{N(\bX_{1};0) \geq n+\sigma_{n},N(\bX_{0};0) = k\right\} \\
&\leq \sum_{k=0}^{n+\tau_{n}} \pr\left\{N(\bX_{1};0) \geq n+\sigma_{n},N(\bX_{0};0) = n+\tau_{n} \right\} \nn \\
&~~~~~~+\sum_{k=n+\tau_{n}}^{2n} \pr\left\{N(\bX_{0};0) = k\right\} \\
\label{ref82}
&\leq (n+\tau_{n}+1) \cdot \pr\left\{N(\bX_{1};0) \geq n+\sigma_{n}|N(\bX_{0};0) = n+\tau_{n} \right\} \nn \\
&~~~~~~+\pr\left\{N(\bX_{0};0) \geq n+\tau_{n}\right\}.
\end{align}
Upper-bounding \eqref{ref81} with \eqref{ref82} yields that
\begin{align}
\pr\{N(\bX_{2};0) = 2n\}
\label{ref83}
&\leq (n+\sigma_{n}+1) \cdot \pr\left\{N(\bX_{2};0) = 2n|N(\bX_{1};0) = n+\sigma_{n} \right\} \nn \\
&~~~~+(n+\tau_{n}+1) \cdot \pr\left\{N(\bX_{1};0) \geq n+\sigma_{n}|N(\bX_{0};0) = n+\tau_{n} \right\} \nn \\
&~~~~+\pr\left\{N(\bX_{0};0) \geq n+\tau_{n}\right\}.
\end{align}

Consider the probability in the first term in \eqref{ref83}. Let us choose $\sigma_{n}=\tfrac{n^{3/4}}{\sqrt{64\lambda}}\sqrt{\log(n^{\rho})}$, $\rho > 0$, and then, according to Proposition \ref{PROP_9}, we get that
\begin{align}
&\pr\left\{N(\bX_{2};0) = 2n|N(\bX_{1};0) = n+\sigma_{n} \right\} \nn \\
&~~~\leq \exp\left\{ -\frac{n \sigma_{n}^{2}}{2(n+\sigma_{n})}
\exp \left\{ - 32\lambda \cdot  \frac{\sigma_{n}^{2}}{\sqrt{n}(n-\sigma_{n})} \right\} \right\} \\
&~~~\leq \exp\left\{ -\frac{n \sigma_{n}^{2}}{2(n+n)}
\exp \left\{ - 32\lambda \cdot  \frac{\sigma_{n}^{2}}{\sqrt{n}(n-\tfrac{n}{2})} \right\} \right\} \\
&~~~= \exp\left\{ -\frac{\sigma_{n}^{2}}{4}
\exp \left\{ - 64\lambda \cdot  \frac{\sigma_{n}^{2}}{n^{3/2}} \right\} \right\} \\
&~~~= \exp\left\{ -\frac{1}{4} \cdot \frac{n^{3/2}}{64\lambda}\log(n^{\rho})
\exp \left\{ -\frac{64\lambda}{n^{3/2}} \cdot \frac{n^{3/2}}{64\lambda}\log(n^{\rho}) \right\} \right\} \\
&~~~= \exp\left\{ -\frac{n^{3/2}}{256\lambda}\log(n^{\rho})
\exp \left\{-\log(n^{\rho}) \right\} \right\} \\
\label{bound1}
&~~~= \exp\left\{ -\frac{\rho}{256\lambda} \log(n) n^{3/2-\rho} \right\},
\end{align}
which converges to zero for any $\rho \in (0,\tfrac{3}{2})$.

Regarding the third term in \eqref{ref83}, let us choose $\tau_{n}=\tfrac{n^{1/2}}{\sqrt{64\lambda}}\sqrt{\log(n^{\kappa})}$, $\kappa > 0$, and then, according to Proposition \ref{PROP_7} with $P_{n}=\tfrac{1}{2}$ and Pinsker's inequality, we get that
\begin{align}
\pr\left\{N(\bX_{0};0) \geq n+\tau_{n}\right\}
&\leq \exp \left\{-2n \cdot D\left(\frac{1}{2} + \frac{\tau_{n}}{2n} \middle\| \frac{1}{2} \right) \right\} \\
&\leq \exp \left\{-4n \cdot \left(\frac{\tau_{n}}{2n}\right)^{2} \right\} \\
&=\exp \left\{-\frac{1}{n} \cdot \frac{n}{64\lambda}\log(n^{\kappa}) \right\} \\
\label{bound2}
&=\exp\left\{-\frac{\kappa}{64\lambda} \log(n) \right\},
\end{align}
which converges to zero as $n \to \infty$.

As for the probability in the second term in \eqref{ref83}, 
\begin{align}
\pr\left\{N(\bX_{1};0) \geq n+\sigma_{n}|N(\bX_{0};0) = n+\tau_{n} \right\}, 
\end{align}
we conclude from Proposition \ref{PROP_8} that for all sufficiently large $n$, the probability to update to `0' is upper-bounded as 
\begin{align} 
Q_{n} 
&\leq \frac{1}{2} + \frac{60\tau_{n}p_{n}}{\lambda n^{1/4-\delta_{n}}} \\
&= \frac{1}{2} + \frac{60}{\lambda n^{1/4}} \cdot
\sqrt{\log(n^{\theta})} \cdot
\frac{\sqrt{n}}{\sqrt{64\lambda}}\sqrt{\log(n^{\kappa})} \cdot \frac{\lambda}{\sqrt{n}}  \\
&= \frac{1}{2} + \frac{60\sqrt{\theta\kappa}\log(n)}{\sqrt{64\lambda} n^{1/4}},  
\end{align} 
and then, it follows from Proposition \ref{PROP_7} that
\begin{align}
&\pr\left\{N(\bX_{1};0) \geq n+\sigma_{n}|N(\bX_{0};0) = n+\tau_{n} \right\} \nn \\
&~~~~\leq \exp \left\{-2n \cdot D\left(\frac{1}{2} + \frac{\sigma_{n}}{2n} \middle\| \frac{1}{2} + \frac{60\sqrt{\theta\kappa}\log(n)}{\sqrt{64\lambda} n^{1/4}} \right) \right\} \\
&~~~~= \exp \left\{-2n \cdot D\left(\frac{1}{2} + \frac{\sqrt{\log(n^{\rho})}}{2\sqrt{64\lambda} n^{1/4}} \middle\| \frac{1}{2} + \frac{60\sqrt{\theta\kappa}\log(n)}{\sqrt{64\lambda} n^{1/4}} \right) \right\} \\
&~~~~\leq \exp \left\{-4n \cdot 
\frac{1}{64\lambda \sqrt{n}}
\left(60\sqrt{\theta\kappa}\log(n)-\frac{1}{2}\sqrt{\log(n^{\rho})}\right)^{2} \right\} \\
\label{bound3}
&~~~~\leq \exp \left\{- \frac{\theta\kappa}{\lambda} \log^{2}(n) \sqrt{n} \right\}.
\end{align}

We now continue from \eqref{ref83}. Using the three bounds from \eqref{bound1}, \eqref{bound2}, and \eqref{bound3} leads to  
\begin{align}
\pr\{N(\bX_{2};0) = 2n\}
&\leq (n+\sigma_{n}+1) \cdot \exp\left\{ -\frac{\rho}{256\lambda} \log(n) n^{3/2-\rho} \right\} \nn \\
&~~~~+(n+\tau_{n}+1) \cdot \exp \left\{- \frac{\theta\kappa}{\lambda} \log^{2}(n) \sqrt{n} \right\} 
+ \exp\left\{-\frac{\kappa}{64\lambda} \log(n) \right\} \\
&\leq 2n \cdot \exp\left\{ -\frac{\rho}{256\lambda} \log(n) n^{3/2-\rho} \right\} \nn \\
&~~~~+2n \cdot \exp \left\{- \frac{\theta\kappa}{\lambda} \log^{2}(n) \sqrt{n} \right\} 
+ \exp\left\{-\frac{\kappa}{64\lambda} \log(n) \right\},
\end{align}
which converges to zero as $n \to \infty$, hence the proof of Theorem \ref{Main_THEOREM2} is complete.

\section*{Appendix A - Proof of Proposition \ref{PROP_1}}
\renewcommand{\theequation}{A.\arabic{equation}}
\setcounter{equation}{0}  

Denote $N_{0}=N(\bX_{1};0)$.  
We would like to prove that the random variable $|N_{0}-n|/\sqrt{n}$ is bounded away from zero with an overwhelmingly high probability at large $n$.
Note that
\begin{align}
N_{0} = \sum_{\ell=1}^{2n} I_{\ell}, 
\end{align}
where $I_{\ell} \sim \text{Ber}(\tfrac{1}{2})$, for all $\ell \in \{1,2,\ldots, 2n\}$, and all of these binary random variables are independent.
Let $\epsilon > 0$ and $\alpha(\epsilon) > 0$, that will be specified later on with the property that $\alpha(\epsilon) \xrightarrow{\epsilon \to 0} 0$.
Consider the following
\begin{align}
\pr \left\{\left|\frac{N_{0}-n}{\sqrt{n}}\right| \geq \alpha(\epsilon) \right\}
\label{ToRef18}
= \pr \left\{\left|\sqrt{2n}\left(\frac{1}{2n}\sum_{\ell=1}^{2n} I_{\ell} - \tfrac{1}{2}\right) \right| \geq \frac{\alpha(\epsilon)}{\sqrt{2}} \right\}.
\end{align}
In order to conclude that the normalized sum inside the probability in \eqref{ToRef18} converge in distribution to normal random variables, we invoke Lindeberg-L\'evy central limit theorem (CLT) \cite[p.\ 144, Theorem 3.4.1.]{DURRETT}. We have the following result.
\begin{theorem}[Lindeberg-L\'evy CLT]
	Suppose $\{X_{n}\}$ is a sequence of IID random variables with $\mathbb{E}[X_{i}]=\mu$ and $\text{Var}[X_{i}]=\sigma^{2}<\infty$. 
	Let us denote $\bar{X}_{n}=\tfrac{1}{n}\sum_{i=1}^{n}X_{i}$.
	Then as $n \to \infty$, the random variables $\sqrt{n}(\bar{X}_{n}-\mu)$ converge in distribution to a normal $\calN(0,\sigma^{2})$.
\end{theorem}    
Now, concerning the normalized sum inside the probability in \eqref{ToRef18}, it follows by Lindeberg-L\'evy CLT that
\begin{align}
\sqrt{2n}\left(\frac{1}{2n}\sum_{\ell=1}^{2n} I_{\ell} - \tfrac{1}{2}\right)
\xrightarrow{d} X \sim \calN(0,\tfrac{1}{4}).
\end{align}   
We continue from \eqref{ToRef18} and arrive at
\begin{align}
\lim_{n \to \infty} \pr \left\{\left|\frac{N_{0}-n}{\sqrt{n}}\right| \geq \alpha(\epsilon) \right\}
&= \pr \left\{\left|X\right| \geq \frac{\alpha(\epsilon)}{\sqrt{2}} \right\} \\
&= \pr \left\{\left|\calN(0,\tfrac{1}{2})\right| \geq \alpha(\epsilon) \right\} \\
&= 1- \frac{\epsilon}{2},
\end{align} 
which can obviously be satisfied by a proper choice of $\alpha(\epsilon)$.
We conclude that for any $\epsilon > 0$, there exists some $M(\epsilon)$, such that for all $n \geq M(\epsilon)$, 
\begin{align}
\pr \left\{\left|\frac{N_{0}-n}{\sqrt{n}}\right| \geq \alpha(\epsilon) \right\}
\geq 1- \epsilon,
\end{align} 
which completes the proof of Proposition \ref{PROP_1}.

\section*{Appendix B - Proof of Proposition \ref{PROP_2}}
\renewcommand{\theequation}{B.\arabic{equation}}
\setcounter{equation}{0}

Assume that the numbers of zeros and ones are $n+\alpha\sqrt{n}$ and $n-\alpha\sqrt{n}$, respectively. 
If an agent starts with a `0', then the probability to decide in favor of `0' is lower-bounded by 
\begin{align}
\label{ToCall0}
Q_{n}^{0}
&=\pr\left\{\text{Bin}\left(n+\alpha\sqrt{n}-1,p_{n}\right) + 1 \geq \text{Bin}\left(n-\alpha\sqrt{n},p_{n}\right) \right\} \\
\label{ref63}
&\geq \pr\left\{\text{Bin}\left(n+\alpha\sqrt{n}-1,p_{n}\right) + \text{Bin}\left(1,p_{n}\right) \geq \text{Bin}\left(n-\alpha\sqrt{n},p_{n}\right) \right\} \\
\label{ref64}
&= \pr\left\{\text{Bin}\left(n,p_{n}\right) + \text{Bin}\left(\alpha\sqrt{n}-1,p_{n}\right) + \text{Bin}\left(1,p_{n}\right) \geq \text{Bin}\left(n-\alpha\sqrt{n},p_{n}\right) \right\} \\ 
\label{ref65}
&\geq \pr\left\{\text{Bin}\left(n,p_{n}\right) + \text{Bin}\left(\alpha \sqrt{n},p_{n}\right) \geq \text{Bin}\left(n,p_{n}\right) \right\} \\
&= \pr\left\{X + Z \geq Y \right\},
\end{align}
where $X,Y \sim \text{Bin}\left(n,p_{n}\right)$ and $Z \sim \text{Bin}\left(\alpha\sqrt{n},p_{n}\right)$ are three independent random variables.
The passage to \eqref{ref63} is true since $\text{Bin}\left(1,p_{n}\right) \leq 1$ with probability 1 and \eqref{ref65} follows by increasing the number of trials in the binomial random variable on the right hand side of the inequality inside the probability in \eqref{ref64}.

It follows from the law of total probability that 
\begin{align}
\pr\left\{X + Z \geq Y \right\}
\label{ref0}
&= \sum_{m=0}^{\alpha \sqrt{n}} \pr\left\{X + m \geq Y \right\} \pr\left\{Z = m \right\}.
\end{align}
As for the probability $\pr\left\{X + m \geq Y \right\}$, we have that 
\begin{align}
\pr\left\{X + m \geq Y \right\} 
\label{ref3}
= \pr\left\{X \geq Y \right\} + \pr\left\{X + 1 = Y \right\} + \ldots + \pr\left\{X + m = Y \right\}.
\end{align}
It follows by symmetry that 
\begin{align}
1 &= \pr\{X > Y\}+\pr\{X < Y\}+\pr\{X = Y\} \\
&= 2\pr\{X > Y\}+\pr\{X = Y\},
\end{align}
or,
\begin{align}
\pr\{X > Y\} = \frac{1}{2} - \frac{1}{2} \cdot \pr\{X = Y\},
\end{align}
which implies that 
\begin{align}
\pr\{X \geq Y\}  
&= \pr\{X > Y\} + \pr\{X = Y\} \\
\label{ref1}
&= \frac{1}{2} + \frac{1}{2} \cdot \pr\{X = Y\}.
\end{align}
Substituting \eqref{ref1} back into \eqref{ref3} yields 
\begin{align}
\pr\left\{X + m \geq Y \right\} 
&= \frac{1}{2} + \frac{1}{2} \cdot \pr\{X = Y\} + \pr\left\{X + 1 = Y \right\} + \ldots + \pr\left\{X + m = Y \right\} \\
\label{Ref0}
&\geq \frac{1}{2} + \frac{1}{2} \cdot \sum_{i=0}^{m} \pr\{X + i = Y \}.
\end{align}
Lower-bounding \eqref{ref0} with \eqref{Ref0} yields
\begin{align}
Q_{n}^{0}
&\geq \sum_{m=0}^{\alpha \sqrt{n}} \pr\left\{X + m \geq Y \right\} \pr\left\{Z = m \right\} \\
&\geq \sum_{m=0}^{\alpha \sqrt{n}} \left[ \frac{1}{2} + \frac{1}{2} \cdot \sum_{i=0}^{m} \pr\{X + i = Y \} \right] \pr\left\{Z = m \right\} \\
\label{ref50}
&= \frac{1}{2} + \frac{1}{2} \cdot \sum_{m=0}^{\alpha \sqrt{n}} \left(\sum_{i=0}^{m} \pr\{X + i = Y \}\right) \pr\left\{Z = m \right\}.
\end{align}
The following result, which is proved in Appendix C, is going to be instrumental.
\begin{lemma} \label{Lemma_Binomial1}
	Let $X,Y \sim \text{Bin}\left(n,p_{n}\right)$ be two independent binomial random variables with $p_{n} = \tfrac{\lambda}{\sqrt{n}}$. Then, for any $i \in \{0,1,\ldots,\lambda \sqrt{n}\}$,
	\begin{align}
	\pr\{X + i = Y\}
	\geq \frac{\pi}{e^{4}}
	\frac{\exp\{-4/\lambda\}}{\lambda n^{1/4}} \exp \left\{ - \frac{i^{2}}{\lambda \sqrt{n}} \right\}.
	\end{align}
\end{lemma}
Recall that $Z \sim \text{Bin}\left(\alpha\sqrt{n},\tfrac{\lambda}{\sqrt{n}}\right)$ and thus
\begin{align}
\mu_{Z} &= \mathbb{E}[Z] = \alpha\lambda, \\
\sigma_{Z} &= \sqrt{\text{Var}[Z]} = \sqrt{\alpha \lambda \left(1 - \frac{\lambda}{\sqrt{n}}\right)} \leq \sqrt{\alpha\lambda}.
\end{align}
Let us define $\nu_{0}=[\alpha\lambda-\sqrt{2\alpha\lambda}]_{+}$ and $\nu_{1}=\alpha\lambda+\sqrt{2\alpha\lambda}$. 
Since $\sqrt{n}$ is assumed to be large, then $\alpha\lambda+\sqrt{2\alpha\lambda} \leq \alpha\sqrt{n}$ is true, and we continue to lower-bound \eqref{ref50} as
\begin{align}
Q_{n}^{0}
&\geq \frac{1}{2} + \frac{1}{2} \cdot \sum_{m=\nu_{0}}^{\nu_{1}} \left(\sum_{i=0}^{m} \pr\{X + i = Y \}\right) \pr\left\{Z = m \right\} \\
&\geq \frac{1}{2} + \sum_{m=\nu_{0}}^{\nu_{1}} \left(\sum_{i=0}^{m} \frac{\pi}{2e^{4}}
\frac{\exp\{-4/\lambda\}}{\lambda n^{1/4}} \exp \left\{ - \frac{i^{2}}{\lambda \sqrt{n}} \right\} \right) \pr\left\{Z = m \right\} \\
&\geq \frac{1}{2} + \sum_{m=\nu_{0}}^{\nu_{1}} \left(\sum_{i=0}^{m} \frac{\pi}{2e^{4}}
\frac{\exp\{-4/\lambda\}}{\lambda n^{1/4}} \exp \left\{ - \frac{m^{2}}{\lambda \sqrt{n}} \right\} \right) \pr\left\{Z = m \right\} \\
&= \frac{1}{2} + \sum_{m=\nu_{0}}^{\nu_{1}} \frac{\pi}{2e^{4}}
\frac{\exp\{-4/\lambda\}(m+1)}{\lambda n^{1/4}} \exp \left\{ - \frac{m^{2}}{\lambda \sqrt{n}} \right\} \pr\left\{Z = m \right\} \\
\label{ref5}
&\geq \frac{1}{2} + \sum_{m=\nu_{0}}^{\nu_{1}} \frac{\pi}{2e^{4}}
\frac{\exp\{-4/\lambda\}(\nu_{0}+1)}{\lambda n^{1/4}} \exp \left\{ - \frac{\nu_{0}^{2}}{\lambda \sqrt{n}} \right\} \pr\left\{Z = m \right\} \\
\label{ref6}
&= \frac{1}{2} + \frac{\pi}{2e^{4}}
\frac{\exp\{-4/\lambda\}(\nu_{0}+1)}{\lambda n^{1/4}} \exp \left\{ - \frac{\nu_{0}^{2}}{\lambda \sqrt{n}} \right\} \pr\left\{\nu_{0} \leq Z \leq \nu_{1} \right\},
\end{align}
where \eqref{ref5} is true since the function $g(t) = (t+1) \exp\left\{-\tfrac{t^{2}}{\lambda \sqrt{n}}\right\}$ is monotonically increasing as long as $t \in \left[0, \tfrac{\sqrt{1+2\lambda \sqrt{n}}-1}{2}\right]$ and since $\sqrt{n}$ is assumed to be large, it then follows that $g(t)$ is monotonically increasing in the entire range $\left[[\alpha\lambda-\sqrt{2\alpha\lambda}]_{+},\alpha\lambda+\sqrt{2\alpha\lambda}\right]$. 

As for the probability in \eqref{ref6}, we split into two cases. If $\nu_{0}>0$, 
\begin{align} 
\pr\left\{\nu_{0} \leq Z \leq \nu_{1} \right\} 
&= \pr\left\{\alpha\lambda-\sqrt{2\alpha\lambda} \leq Z \leq \alpha\lambda+\sqrt{2\alpha\lambda} \right\} \\
&\geq \pr\left\{\mu_{Z} - \sqrt{2}\sigma_{Z} \leq Z \leq \mu_{Z} + \sqrt{2}\sigma_{Z} \right\} \\
&= \pr\left\{ |Z - \mu_{Z}| \leq \sqrt{2}\sigma_{Z} \right\} \\
\label{ref7}
&\geq \frac{1}{2},
\end{align}
where \eqref{ref7} is due to Chebyshev's inequality.
Otherwise, if $\nu_{0} = 0$, which is equivalent to $\alpha\lambda\leq \sqrt{2\alpha\lambda}$, we have that
\begin{align}
\pr\left\{Z \geq \nu_{1}\right\}
&\leq \frac{\mathbb{E}[Z]}{\nu_{1}} \\
&= \frac{\alpha\lambda}{\alpha\lambda + \sqrt{2\alpha\lambda}} \\
&\leq \frac{\alpha\lambda}{\alpha\lambda + \alpha\lambda} \\
&= \frac{1}{2},
\end{align}
and then,
\begin{align} 
\pr\left\{\nu_{0} \leq Z \leq \nu_{1} \right\} 
&= \pr\left\{0 \leq Z \leq \nu_{1} \right\} \\
&= 1 - \pr\left\{Z \geq \nu_{1} \right\} \\
\label{ref8}
&\geq \frac{1}{2}.
\end{align}

Lower-bounding the probability in \eqref{ref6} by $\tfrac{1}{2}$, we finally arrive at 
\begin{align}
Q_{n}^{0}
&\geq \frac{1}{2} + \frac{\pi}{4e^{4}}
\frac{([\alpha\lambda-\sqrt{2\alpha\lambda}]_{+}+1)\exp\{-4/\lambda\}}{\lambda n^{1/4}} \exp \left\{ - \frac{([\alpha\lambda-\sqrt{2\alpha\lambda}]_{+})^{2}}{\lambda \sqrt{n}} \right\} \\ 
&\geq \frac{1}{2} + \frac{\pi}{8e^{4}}
\frac{([\alpha\lambda-\sqrt{2\alpha\lambda}]_{+}+1)\exp\{-4/\lambda\}}{\lambda n^{1/4}},
\end{align}
where the last inequality is, again, due to the fact that $\sqrt{n}$ is assumed to be large. The proof of Proposition \ref{PROP_2} is now complete.

\section*{Appendix C - Proof of Lemma \ref{Lemma_Binomial1}}
\renewcommand{\theequation}{C.\arabic{equation}}
\setcounter{equation}{0}

The probability $\pr\{X + i = Y \}$ can be written explicitly as 
\begin{align}
\pr\{X + i = Y \} = \sum_{\ell=0}^{n} \sum_{k=0}^{n}
\binom{n}{\ell} p_{n}^{\ell} (1-p_{n})^{n-\ell}
\binom{n}{k} p_{n}^{k} (1-p_{n})^{n-k} \IND \{\ell + i = k\},
\end{align}
which implies that
\begin{align}
\label{ToCont0}
\pr\{X + i = Y \} \geq \sum_{\ell=1}^{n-1} \sum_{k=1}^{n-1}
\binom{n}{\ell} p_{n}^{\ell} (1-p_{n})^{n-\ell}
\binom{n}{k} p_{n}^{k} (1-p_{n})^{n-k} \IND \{\ell + i = k\}.
\end{align} 
We continue by lower-bounding the PMF of the binomial random variable $X=\text{Bin}(n,p)$, which is given by
\begin{align}
\label{TermToCall4}
P_{X}(k) = \binom{n}{k} p^{k} (1-p)^{n-k},~~~k \in [0:n].
\end{align}
In order to lower-bound the binomial coefficient in \eqref{TermToCall4}, we use the Stirling's bounds in
\begin{align}
\label{Stirling}
\sqrt{2 \pi n} \cdot n^{n} \cdot e^{-n} 
\leq n! 
\leq e \sqrt{n} \cdot n^{n} \cdot e^{-n},~~n \geq 1, 
\end{align} 
and get that
\begin{align}
\binom{n}{k}
&= \frac{n!}{k! \cdot (n-k)!} \\
\label{TermToCall5}
&\geq \frac{\sqrt{2\pi}}{e^{2}} \sqrt{\frac{n}{k(n-k)}}
\exp \left\{ -n \left[\frac{k}{n} \log \left(\frac{k}{n}\right) 
+ \left(1-\frac{k}{n}\right) \log \left(1-\frac{k}{n}\right) \right] \right\}.
\end{align}
Substituting \eqref{TermToCall5} back into \eqref{TermToCall4} yields that for any $k=1,2,\ldots,n-1$
\begin{align}
P_{X}(k) 
&\geq \frac{\sqrt{2\pi}}{e^{2}} \sqrt{\frac{n}{k(n-k)}}
\exp \left\{ -n D\left(\frac{k}{n} \middle\| p \right) \right\} \\
\label{ToRef4}
&\geq \frac{\sqrt{2\pi}}{e^{2}} \sqrt{\frac{1}{k}}
\exp \left\{ -n D\left(\frac{k}{n} \middle\| p \right) \right\},
\end{align}
where $D(\alpha \| \beta)$, for $\alpha,\beta \in [0,1]$, is defined in \eqref{DEF_Bin_DIVERGENCE}.
Substituting twice the lower bound of \eqref{ToRef4} into \eqref{ToCont0}, we arrive at  
\begin{align}
\pr\{X + i = Y \} 
&\geq 
\frac{2\pi}{e^{4}}
\sum_{\ell=1}^{n-1} \sum_{k=1}^{n-1}
\sqrt{\frac{1}{\ell k}}
\exp \left\{ -n D\left(\frac{\ell}{n} \middle\| p_{n} \right) \right\} \nn \\
&~~~~~~~~~~~~~~~~~~~~~\times 
\exp \left\{ -n D\left(\frac{k}{n} \middle\| p_{n} \right) \right\} \IND \{\ell + i = k\} \\
&= 
\frac{2\pi}{e^{4}}
\sum_{\ell=1}^{n-i-1}
\sqrt{\frac{1}{\ell (\ell+i)}}
\exp \left\{ -n D\left(\frac{\ell}{n} \middle\| p_{n} \right) \right\} \cdot 
\exp \left\{ -n D\left(\frac{\ell + i}{n} \middle\| p_{n} \right) \right\} \\
\label{ToRef1}
&\geq 
\frac{2\pi}{e^{4}}
\sum_{\ell=1}^{n-i-1}
\frac{1}{\ell+i}
\exp \left\{ -n D\left(\frac{\ell}{n} \middle\| p_{n} \right) \right\} \cdot 
\exp \left\{ -n D\left(\frac{\ell + i}{n} \middle\| p_{n} \right) \right\}.
\end{align}
In order to lower-bound \eqref{ToRef1}, 
let $\epsilon_{n} = \tfrac{1}{n^{3/4}}$, for $n=1,2, \ldots$ and define the set of numbers
\begin{align}
\calN_{n} = \left\{n(p_{n}-\epsilon_{n})-\tfrac{i}{2}+1,n(p_{n}-\epsilon_{n})-\tfrac{i}{2}+2,\ldots,np_{n}-\tfrac{i}{2},\ldots,n(p_{n}+\epsilon_{n})-\tfrac{i}{2}\right\},
\end{align}
whose cardinality is given by
\begin{align}
\label{Cardinality}
|\calN_{n}| = 2 n \epsilon_{n}.
\end{align}
We now continue from \eqref{ToRef1} and arrive at
\begin{align}
\pr\{X + i = Y\}
&\geq \frac{2\pi}{e^{4}}
\sum_{\ell=1}^{n-i-1}
\frac{1}{\ell+i}
\exp \left\{ -n D\left(\frac{\ell}{n} \middle\| p_{n} \right) \right\} \cdot 
\exp \left\{ -n D\left(\frac{\ell + i}{n} \middle\| p_{n} \right) \right\} \\
\label{ToRef2}
&\geq \frac{2\pi}{e^{4}}
\sum_{\ell \in \calN_{n}}
\frac{1}{\ell+i}
\exp \left\{ -n D\left(\frac{\ell}{n} \middle\| p_{n} \right) \right\} \cdot 
\exp \left\{ -n D\left(\frac{\ell + i}{n} \middle\| p_{n} \right) \right\}.
\end{align}
We upper-bound the exponents in \eqref{ToRef2} using a reversed Pinsker inequality. Recall that the total variation distance between two probability distributions $P$ and $Q$ is defined by
\begin{align} \label{DEF_TVD}
|P-Q| = \frac{1}{2} \sum_{x \in \calX} |P(x)-Q(x)|,
\end{align}   
and the Kullback-Leibler divergence is defined by
\begin{align}
D(P\|Q) = \sum_{x \in \calX} P(x) \log \frac{P(x)}{Q(x)}.
\end{align}
Then, it holds that \cite[p.\ 5974, Eq.\ (23)]{SASON}
\begin{align} \label{Reverse_PINSKER}
D(P\|Q) \leq \left(\frac{2}{Q_{\mbox{\tiny min}}}\right) \cdot |P-Q|^{2},
\end{align}  
when
\begin{align}
Q_{\mbox{\tiny min}} = \min_{x\in \calX} Q(x).
\end{align}
Now, the exponent in \eqref{ToRef2} is lower-bounded by
\begin{align}
\pr\{X + i = Y\}
&\geq \frac{2\pi}{e^{4}}
\sum_{\ell \in \calN_{n}}
\frac{1}{\ell+i}
\exp \left\{ -n \cdot \frac{2}{p_{n}} \cdot \left(\frac{\ell}{n} - p_{n} \right)^{2} \right\} \cdot 
\exp \left\{ -n \cdot \frac{2}{p_{n}} \cdot \left(\frac{\ell + i}{n} - p_{n} \right)^{2} \right\} \\
&= \frac{2\pi}{e^{4}}
\sum_{\ell \in \calN_{n}}
\frac{1}{\ell+i}
\exp \left\{ - \frac{2}{np_{n}} \cdot \left(\ell - np_{n} \right)^{2} \right\} \cdot 
\exp \left\{ -\frac{2}{np_{n}} \cdot \left(\ell + i - np_{n} \right)^{2} \right\} \\
&= \frac{2\pi}{e^{4}}
\sum_{\ell \in \calN_{n}}
\frac{1}{\ell+i}
\exp \left\{ - \frac{2}{np_{n}} \cdot \left[(\ell - np_{n})^{2} + (\ell + i - np_{n} )^{2} \right] \right\} \\
&= \frac{2\pi}{e^{4}}
\sum_{\ell \in \calN_{n}}
\frac{1}{\ell+i}
\exp \left\{ - \frac{2}{np_{n}} \cdot \left[2\left(\ell + \frac{i}{2} - np_{n}\right)^{2} + \frac{i^{2}}{2} \right] \right\}.
\end{align}
Both the term $\ell+i$ and the exponent are maximized at $\ell = n(p_{n}+\epsilon_{n})-\tfrac{i}{2}$, and thus
\begin{align}
\pr\{X + i = Y\}
&\geq \frac{2\pi}{e^{4}}
\sum_{\ell \in \calN_{n}}
\frac{1}{n(p_{n}+\epsilon_{n})+\tfrac{i}{2}}
\exp \left\{ - \frac{2}{np_{n}} \cdot \left( 2n^{2}\epsilon_{n}^{2} + \frac{i^{2}}{2} \right) \right\} \\
&= \frac{2\pi}{e^{4}}
\frac{|\calN_{n}|}{n(p_{n}+\epsilon_{n})+\tfrac{i}{2}}
\exp \left\{ - \frac{2}{np_{n}} \cdot \left( 2n^{2}\epsilon_{n}^{2} + \frac{i^{2}}{2} \right) \right\} \\
&= \frac{2\pi}{e^{4}}
\frac{2n\epsilon_{n}}{n(p_{n}+\epsilon_{n})+ \tfrac{i}{2}} \exp \left\{ - \left( \frac{4n\epsilon_{n}^{2}}{p_{n}} + \frac{i^{2}}{np_{n}} \right) \right\}.
\end{align}
Note that
\begin{align}
4n \cdot \frac{1}{p_{n}} \cdot \epsilon_{n}^{2} 
&= 4n \cdot \frac{\sqrt{n}}{\lambda} \cdot \left(\frac{1}{n^{3/4}}\right)^{2}  \\
&= \frac{4}{\lambda}.
\end{align}
In addition, $i \leq n(p_{n}+\epsilon_{n})$, since we assume that $i \in \{0,1,\ldots,\lambda \sqrt{n}\}$, and thus,
\begin{align}
\pr\{X + i = Y\}
&\geq \frac{2\pi}{e^{4}}
\frac{2n\epsilon_{n}}{n(p_{n}+\epsilon_{n})+ n(p_{n}+\epsilon_{n})} \exp \left\{ - \left( \frac{4}{\lambda} + \frac{i^{2}}{np_{n}} \right) \right\} \\
\label{ToRef3}
&= \frac{2\pi}{e^{4}}
\frac{\epsilon_{n}}{p_{n}+\epsilon_{n}} \exp \left\{ - \left( \frac{4}{\lambda} + \frac{i^{2}}{np_{n}} \right) \right\}.
\end{align}
As for the fraction $\epsilon_{n}/(p_{n}+\epsilon_{n})$,
\begin{align}
\frac{\epsilon_{n}}{p_{n}+\epsilon_{n}}  
&= \frac{\frac{1}{n^{3/4}}}{\frac{\lambda}{\sqrt{n}}+\frac{1}{n^{3/4}}}  \\
&\geq  \frac{\frac{1}{n^{3/4}}}{\frac{\lambda}{\sqrt{n}}+\frac{\lambda}{\sqrt{n}}}  \\
&= \frac{\sqrt{n}}{2\lambda n^{3/4}}  \\
&= \frac{1}{2 \lambda n^{1/4}},
\end{align}
and substituting it back into \eqref{ToRef3} yields
\begin{align}
\pr\{X + i = Y\}
&\geq \frac{\pi}{e^{4}}
\frac{1}{\lambda n^{1/4}} \exp \left\{ - \left( \frac{4}{\lambda} + \frac{i^{2}}{\lambda \sqrt{n}} \right) \right\} \\
&= \frac{\pi}{e^{4}}
\frac{\exp\{-4/\lambda\}}{\lambda n^{1/4}} \exp \left\{ - \frac{i^{2}}{\lambda \sqrt{n}} \right\}.
\end{align}
The proof of Lemma \ref{Lemma_Binomial1} is complete.

\section*{Appendix D - Proof of Proposition \ref{PROP_3}}
\renewcommand{\theequation}{D.\arabic{equation}}
\setcounter{equation}{0}

Let us denote   
\begin{align}
\label{DEF_PHI_EPSILON}
\phi_{n} = \frac{1}{2} + 
\frac{C_{0}(\alpha,\lambda)}{n^{1/4}},~~~\epsilon_{n}=\frac{C_{0}(\alpha,\lambda)}{2n^{1/4}},
\end{align}
and let $Q_{n}^{0}, Q_{n}^{1}$ denote the probabilities of deciding `0', for the two possible initial states. 
It follows from Proposition \ref{PROP_2} that $\min\{Q_{n}^{0},Q_{n}^{1}\} \geq \phi_{n}$.  

We now prove that the probability of drawing a relatively small number of zeros tends to 0 as $n \to \infty$. Denote $N_{0}=N(\bX_{1};0)$ and consider the following for $s \geq 0$
\begin{align}
\pr \left\{N_{0} \leq 2n (\phi_{n}-\epsilon_{n}) \right\}
&= \pr \left\{e^{-sN_{0}} \geq e^{-2ns (\phi_{n}-\epsilon_{n})} \right\} \\
\label{B_ToExp3}
&\leq \frac{\mathbb{E} \left[e^{-sN_{0}}\right]}{e^{-2ns (\phi_{n}-\epsilon_{n})}},
\end{align}   
where \eqref{B_ToExp3} is due to Markov's inequality. Since \eqref{B_ToExp3} holds for every $s \geq 0$, it follows that
\begin{align}
\label{ToRef15}
\pr \left\{N_{0} \leq 2n (\phi_{n}-\epsilon_{n}) \right\}
\leq \inf_{s > 0} \frac{\mathbb{E} \left[e^{-sN_{0}}\right]}{e^{-2ns (\phi_{n}-\epsilon_{n})}}.
\end{align}
Note that
\begin{align}
N_{0} = \sum_{\ell=1}^{n+\alpha\sqrt{n}} I_{\ell} + \sum_{k=1}^{n-\alpha\sqrt{n}} J_{k},
\end{align}
where $I_{\ell} \sim \text{Ber}(Q_{n}^{0})$, for all $\ell \in \{1,2,\ldots, n+\alpha\sqrt{n}\}$, $J_{k} \sim \text{Ber}(Q_{n}^{1})$, for all $k \in \{1,2,\ldots, n-\alpha\sqrt{n}\}$, and all of these binary random variables are independent.
We get that
\begin{align}
\mathbb{E}\left[e^{-s N_{0}}\right]
&= \mathbb{E}\left[\exp\left\{-s \left(\sum_{\ell=1}^{n+\alpha\sqrt{n}} I_{\ell} + \sum_{k=1}^{n-\alpha\sqrt{n}} J_{k}\right) \right\}\right] \\
&= \mathbb{E}\left[\prod_{\ell=1}^{n+\alpha\sqrt{n}} e^{-s I_{\ell}} \cdot \prod_{k=1}^{n-\alpha\sqrt{n}} e^{-s J_{k}} \right] \\
\label{B_ToExp4}
&= \prod_{\ell=1}^{n+\alpha\sqrt{n}} \mathbb{E}\left[ e^{-s I_{\ell}} \right] \cdot \prod_{k=1}^{n-\alpha\sqrt{n}} \mathbb{E}\left[ e^{-s J_{k}} \right] \\
&= \left[1+Q_{n}^{0}(e^{-s}-1)\right]^{n+\alpha\sqrt{n}} \cdot \left[1+Q_{n}^{1}(e^{-s}-1)\right]^{n-\alpha\sqrt{n}}  \\
\label{B_ToExp5}
&\leq \left[1+\phi_{n}(e^{-s}-1)\right]^{n+\alpha\sqrt{n}} \cdot \left[1+\phi_{n}(e^{-s}-1)\right]^{n-\alpha\sqrt{n}}  \\
\label{ToRef14}
&= \left[1+\phi_{n}(e^{-s}-1)\right]^{2n},
\end{align}
where \eqref{B_ToExp4} is due to the independence of all binary random variables and \eqref{B_ToExp5} is true since $\min\{Q_{n}^{0},Q_{n}^{1}\} \geq \phi_{n}$ and $e^{-s}-1 \leq 0$.  
Substituting \eqref{ToRef14} back into \eqref{ToRef15} yields that
\begin{align}
\pr \left\{N_{0} \leq 2n (\phi_{n}-\epsilon_{n}) \right\}
&\leq \inf_{s > 0} \exp \left\{2n \log \left[1+\phi_{n}(e^{-s}-1)\right] + 2ns (\phi_{n}-\epsilon_{n})\right\} \\
\label{ToRef16}
&= \exp \left\{ 2n \cdot \inf_{s > 0} \{\log \left[1+\phi_{n}(e^{-s}-1)\right] + s (\phi_{n}-\epsilon_{n}) \}\right\}.
\end{align}
Upon defining 
\begin{align}
\label{ToRef17}
g(s) = \log \left[1+\phi_{n}(e^{-s}-1)\right] + s (\phi_{n}-\epsilon_{n}),
\end{align}
we find that the solution to $g'(s)=0$ is given by 
\begin{align}
s^{*} = \log \left(\frac{\phi_{n}[1-(\phi_{n}-\epsilon_{n})]}{(1-\phi_{n})(\phi_{n}-\epsilon_{n})}\right).
\end{align}
Substituting it back into \eqref{ToRef17} yields that
\begin{align}
g(s^{*}) 
&= \log\left(1+ \phi_{n}\left[\frac{(1-\phi_{n})(\phi_{n}-\epsilon_{n})}{\phi_{n}[1-(\phi_{n}-\epsilon_{n})]}-1\right]\right) \nn \\
&~~~~~~~~~~~~~+ (\phi_{n}-\epsilon_{n}) \log \left(\frac{\phi_{n}[1-(\phi_{n}-\epsilon_{n})]}{(1-\phi_{n})(\phi_{n}-\epsilon_{n})}\right) \\
&= \log \left(\frac{1-\phi_{n}}{1-(\phi_{n}-\epsilon_{n})}\right) 
+(\phi_{n}-\epsilon_{n}) \log \left(\frac{\phi_{n}}{\phi_{n}-\epsilon_{n}}\right) \nn \\
&~~~~~~~~~~~~~+ (\phi_{n}-\epsilon_{n}) \log \left(\frac{1-(\phi_{n}-\epsilon_{n})}{1-\phi_{n}}\right)  \\
&= - (\phi_{n}-\epsilon_{n}) \log \left(\frac{\phi_{n}-\epsilon_{n}}{\phi_{n}}\right)
- (1-(\phi_{n}-\epsilon_{n})) \log \left(\frac{1-(\phi_{n}-\epsilon_{n})}{1-\phi_{n}}\right) \\
\label{ToDoPinsker}
&= -D(\phi_{n}-\epsilon_{n} \| \phi_{n}).
\end{align}
We upper-bound the expression in \eqref{ToDoPinsker} using Pinsker's inequality \cite{C1967,Kullback}, which asserts that  
\begin{align} \label{PINSKER}
D(P\|Q) \geq 2 |P-Q|^{2}.
\end{align}
Thus, we arrive at 
\begin{align}
\pr \left\{N_{0} \leq 2n (\phi_{n}-\epsilon_{n}) \right\}
&\leq \exp \left\{ -2n D(\phi_{n}-\epsilon_{n} \| \phi_{n}) \right\} \\
&\leq \exp \left\{ -4n \epsilon_{n}^{2} \right\}.
\end{align}
Hence, we conclude that 
\begin{align}
\pr \left\{N_{0} \geq 2n (\phi_{n}-\epsilon_{n}) \right\}
&\geq 1 - \exp\left\{-4n\epsilon_{n}^{2} \right\},
\end{align} 
and by substituting the specific expressions of $\phi_{n}$ and $\epsilon_{n}$ from \eqref{DEF_PHI_EPSILON}, we arrive at 
\begin{align}
\pr \left\{N_{0} \geq 2n \left(\frac{1}{2} + 
\frac{C_{0}(\alpha,\lambda)}{n^{1/4}}-\frac{C_{0}(\alpha,\lambda)}{2n^{1/4}}\right) \right\}
&\geq 1 - \exp\left\{-4n \left(\frac{C_{0}(\alpha,\lambda)}{2n^{1/4}}\right)^{2} \right\},
\end{align}
or
\begin{align}
\pr \left\{N_{0} \geq  n + C_{0}(\alpha,\lambda) n^{3/4} \right\}
&\geq 1 - \exp\left\{- C_{0}(\alpha,\lambda)^{2} \sqrt{n} \right\},
\end{align}
which converges to 1 as $n \to \infty$. 
Proposition \ref{PROP_3} is now proved.

\section*{Appendix E - Proof of Proposition \ref{PROP_4}}
\renewcommand{\theequation}{E.\arabic{equation}}
\setcounter{equation}{0}

Assume that the numbers of zeros and ones are $n+\beta n^{\frac{3}{4}}$ and $n-\beta n^{\frac{3}{4}}$, respectively. 
If an agent starts with a `0', then the probability to decide in favor of `0' is lower-bounded by 
\begin{align}
Q_{n}^{0}
&=\pr\left\{\text{Bin}\left(n+\beta n^{\frac{3}{4}}-1,p_{n}\right) + 1 \geq \text{Bin}\left(n-\beta n^{\frac{3}{4}},p_{n}\right) \right\} \\
&\geq \pr\left\{\text{Bin}\left(n+\beta n^{\frac{3}{4}}-1,p_{n}\right) + \text{Bin}\left(1,p_{n}\right) \geq \text{Bin}\left(n-\beta n^{\frac{3}{4}},p_{n}\right) \right\} \\ 
&\geq \pr\left\{\text{Bin}\left(n,p_{n}\right) + \text{Bin}\left(\beta n^{\frac{3}{4}},p_{n}\right) \geq \text{Bin}\left(n,p_{n}\right) \right\} \\
&= \pr\left\{X + Z \geq Y \right\},
\end{align}
where $X,Y \sim \text{Bin}\left(n,p_{n}\right)$ and $Z \sim \text{Bin}\left(\beta n^{\frac{3}{4}},p_{n}\right)$.
It follows from the law of total probability that 
\begin{align}
\pr\left\{X + Z \geq Y \right\}
\label{ref10}
&= \sum_{m=0}^{\beta n^{\frac{3}{4}}} \pr\left\{X + m \geq Y \right\} \pr\left\{Z = m \right\}.
\end{align}
As for the probability $\pr\left\{X + m \geq Y \right\}$, recall from \eqref{ref0} that 
\begin{align} 
\pr\left\{X + m \geq Y \right\} 
\label{ref11}
&\geq \frac{1}{2} + \frac{1}{2} \cdot \sum_{i=0}^{m} \pr\{X + i = Y \}.
\end{align}
Lower-bounding \eqref{ref10} with \eqref{ref11} provides
\begin{align}
Q_{n}^{0}
&\geq \sum_{m=0}^{\beta n^{\frac{3}{4}}} \pr\left\{X + m \geq Y \right\} \pr\left\{Z = m \right\} \\
&\geq \sum_{m=0}^{\beta n^{\frac{3}{4}}} \left[ \frac{1}{2} + \frac{1}{2} \cdot \sum_{i=0}^{m} \pr\{X + i = Y \} \right] \pr\left\{Z = m \right\} \\
\label{ref13}
&= \frac{1}{2} + \frac{1}{2} \cdot \sum_{m=0}^{\beta n^{\frac{3}{4}}} \sum_{i=0}^{m} \pr\{X + i = Y \} \pr\left\{Z = m \right\}.
\end{align}
Recall that $Z \sim \text{Bin}\left(\beta n^{\frac{3}{4}},\tfrac{\lambda}{\sqrt{n}}\right)$ and thus, it follows from \eqref{ToRef4} that
\begin{align} \label{Binomial_bound}
P_{Z}(m) \geq \frac{\sqrt{2\pi}}{e^{2}} \sqrt{\frac{1}{m}}
\exp \left\{ -\beta n^{\frac{3}{4}} D\left(\frac{m}{\beta n^{\frac{3}{4}}} \middle\| \frac{\lambda}{\sqrt{n}} \right) \right\}.
\end{align}
In order to lower-bound \eqref{ref13}, 
let $\delta_{n}$, $n=1,2, \ldots$, which converges to zero faster than $p_{n}=\tfrac{\lambda}{\sqrt{n}}$ and define the set of numbers
\begin{align}
\calS_{n} = \left\{\beta n^{\frac{3}{4}}(p_{n}-\delta_{n}),\beta n^{\frac{3}{4}}(p_{n}-\delta_{n})+1,\ldots,\beta n^{\frac{3}{4}}p_{n},\ldots,\beta n^{\frac{3}{4}}(p_{n}+\delta_{n})\right\},
\end{align}
whose cardinality is given by
\begin{align}
\label{Cardinality2}
|\calS_{n}| = 2 \beta n^{\frac{3}{4}} \delta_{n} + 1.
\end{align}
Continuing from \eqref{ref13},
\begin{align}
Q_{n}^{0}
&\geq \frac{1}{2} + \frac{1}{2} \cdot \sum_{m \in \calS_{n}} \sum_{i=0}^{m} \pr\{X + i = Y \} \pr\left\{Z = m \right\} \\
\label{ref18}
&\geq \frac{1}{2} + \frac{1}{2} \cdot \sum_{m \in \calS_{n}} \sum_{i=0}^{m} \frac{\pi}{e^{4}}
\frac{\exp\{-4/\lambda\}}{\lambda n^{1/4}} \exp \left\{ - \frac{i^{2}}{\lambda \sqrt{n}} \right\} \pr\left\{Z = m \right\} \\
&\geq \frac{1}{2} + \sum_{m \in \calS_{n}} \sum_{i=0}^{m} \frac{\pi}{2e^{4}}
\frac{\exp\{-4/\lambda\}}{\lambda n^{1/4}} \exp \left\{ - \frac{m^{2}}{\lambda \sqrt{n}} \right\} \pr\left\{Z = m \right\} \\
\label{ref19}
&\geq \frac{1}{2} + \sum_{m \in \calS_{n}}  \frac{\pi}{2e^{4}}
\frac{m\exp\{-4/\lambda\}}{\lambda n^{1/4}} \exp \left\{ - \frac{m^{2}}{\lambda \sqrt{n}} \right\} \pr\left\{Z = m \right\},
\end{align}
where \eqref{ref18} is due to Lemma \ref{Lemma_Binomial1} (Appendix B) and \eqref{ref19} is because we lower-bounded $m+1$ by $m$. 
Lower-bounding \eqref{ref19} with \eqref{Binomial_bound} yields that
\begin{align}
Q_{n}^{0}
&\geq \frac{1}{2} + \sum_{m \in \calS_{n}} \frac{\pi}{2e^{4}}
\frac{m\exp\{-4/\lambda\}}{\lambda n^{1/4}} \exp \left\{ - \frac{m^{2}}{\lambda \sqrt{n}} \right\} \cdot \frac{\sqrt{2\pi}}{e^{2}} \sqrt{\frac{1}{m}} 
\exp \left\{ -\beta n^{\frac{3}{4}} D\left(\frac{m}{\beta n^{\frac{3}{4}}} \middle\| \frac{\lambda}{\sqrt{n}} \right) \right\} \\
\label{ref14}
&= \frac{1}{2} + \sum_{m \in \calS_{n}} \frac{\pi \sqrt{2\pi}\exp\{-4/\lambda\}}{2e^{6}\lambda}
\frac{\sqrt{m}}{n^{1/4}} \exp \left\{ - \frac{m^{2}}{\lambda \sqrt{n}} \right\} \cdot \exp \left\{ -\beta n^{\frac{3}{4}} D\left(\frac{m}{\beta n^{\frac{3}{4}}} \middle\| \frac{\lambda}{\sqrt{n}} \right) \right\}.
\end{align}
Since $m \in \calS_{n}$, the factor $\tfrac{\sqrt{m}}{n^{1/4}}$ is lower-bounded by
\begin{align}
\frac{\sqrt{m}}{n^{1/4}}
&\geq \frac{\sqrt{\beta n^{\frac{3}{4}}(p_{n}-\delta_{n})}}{n^{1/4}} \\
\label{ref20}
&\geq \frac{\sqrt{\beta n^{\frac{3}{4}}(p_{n}-\tfrac{1}{2}p_{n})}}{n^{1/4}} \\
&= \frac{\sqrt{\tfrac{1}{2}\beta n^{\frac{3}{4}}p_{n}}}{n^{1/4}} \\
&= \frac{\sqrt{\tfrac{1}{2}\beta n^{\frac{3}{4}}\frac{\lambda}{\sqrt{n}}}}{n^{1/4}} \\
\label{ref15}
&= \sqrt{\tfrac{1}{2}\beta\lambda} n^{-\frac{1}{8}},
\end{align}
where \eqref{ref20} holds for all $n$ sufficiently large, since $\delta_{n}$ converges to zero faster than $p_{n}$.
For the first exponent in \eqref{ref14}, it attains its minimal value for the maximal value of $m$ in $\calS_{n}$:
\begin{align}
\exp \left\{ -\frac{m^{2}}{\lambda \sqrt{n}} \right\}
&\geq \exp \left\{ - \frac{\left[\beta n^{\frac{3}{4}}(p_{n}+\delta_{n})\right]^{2}}{\lambda \sqrt{n}} \right\} \\
&= \exp \left\{-\frac{\beta^{2} n^{\frac{3}{2}}(p_{n}+\delta_{n})^{2}}{\lambda \sqrt{n}} \right\} \\
&= \exp \left\{-\frac{\beta^{2} }{\lambda} n (p_{n}+\delta_{n})^{2} \right\} \\
\label{ref21}
&\geq \exp \left\{-\frac{\beta^{2} }{\lambda} n (p_{n}+p_{n})^{2} \right\} \\
&= \exp \left\{-\frac{\beta^{2} }{\lambda} n \frac{4\lambda^{2}}{n} \right\} \\
\label{ref16}
&= \exp \left\{-4\beta^{2}\lambda \right\},
\end{align}
where \eqref{ref21} holds for all $n$ sufficiently large, since $\delta_{n}$ converges to zero faster than $p_{n}$.

For the second exponent in \eqref{ref14}, since we will use the reverse Pinsker inequality in order to upper-bound the binary divergence, it follows that choosing either of the endpoints of $\calS_{n}$ will yield the minimal value. We get that   
\begin{align}
\exp \left\{ -\beta n^{\frac{3}{4}} D\left(\frac{m}{\beta n^{\frac{3}{4}}} \middle\| \frac{\lambda}{\sqrt{n}} \right) \right\}
&\geq \exp \left\{ -\beta n^{\frac{3}{4}} D\left(\frac{\beta n^{\frac{3}{4}}(p_{n}+\delta_{n})}{\beta n^{\frac{3}{4}}} \middle\| \frac{\lambda}{\sqrt{n}} \right) \right\} \\
&= \exp \left\{ -\beta n^{\frac{3}{4}} D\left(\frac{\lambda}{\sqrt{n}}+\delta_{n} \middle\| \frac{\lambda}{\sqrt{n}} \right) \right\} \\
\label{ref22}
&\geq \exp \left\{ -\beta n^{\frac{3}{4}} \frac{2\sqrt{n}}{\lambda} \delta_{n}^{2} \right\} \\
\label{ref17}
&= \exp \left\{ -\frac{2\beta}{\lambda} n^{\frac{5}{4}} \delta_{n}^{2} \right\},
\end{align} 
where \eqref{ref22} follows from the reverse Pinsker inequality in \eqref{Reverse_PINSKER}.
Lower-bounding \eqref{ref14} using \eqref{ref15}, \eqref{ref16}, and \eqref{ref17} yields that 
\begin{align}
Q_{n}^{0}
&\geq \frac{1}{2} + \sum_{m \in \calS_{n}} \frac{\pi \sqrt{2\pi}\exp\{-4/\lambda\}}{2e^{6}\lambda}
\sqrt{\tfrac{1}{2}\beta\lambda} n^{-\frac{1}{8}} \cdot \exp \left\{-4\beta^{2}\lambda \right\} \cdot \exp \left\{ -\frac{2\beta}{\lambda} n^{\frac{5}{4}} \delta_{n}^{2} \right\} \\
&= \frac{1}{2} + \sum_{m \in \calS_{n}} \frac{\pi^{\frac{3}{2}} \exp\{-4/\lambda\} \sqrt{\beta}\exp \left\{-4\beta^{2}\lambda \right\}}{2e^{6}\sqrt{\lambda}} n^{-\frac{1}{8}} \cdot \exp \left\{ -\frac{2\beta}{\lambda} n^{\frac{5}{4}} \delta_{n}^{2} \right\} \\
&= \frac{1}{2} + \frac{\pi^{\frac{3}{2}} \exp\{-4/\lambda\} \sqrt{\beta}\exp \left\{-4\beta^{2}\lambda \right\}}{2e^{6}\sqrt{\lambda}} \cdot |\calS_{n}| \cdot n^{-\frac{1}{8}} \cdot \exp \left\{ -\frac{2\beta}{\lambda} n^{\frac{5}{4}} \delta_{n}^{2} \right\} \\
&\geq \frac{1}{2} + \frac{\pi^{\frac{3}{2}} \exp\{-4/\lambda\} \sqrt{\beta}\exp \left\{-4\beta^{2}\lambda \right\}}{2e^{6}\sqrt{\lambda}} \cdot 2 \beta n^{\frac{3}{4}} \delta_{n} \cdot n^{-\frac{1}{8}} \cdot \exp \left\{ -\frac{2\beta}{\lambda} n^{\frac{5}{4}} \delta_{n}^{2} \right\} \\
&= \frac{1}{2} + \frac{(\pi\beta)^{\frac{3}{2}} \exp\{-4/\lambda\} \exp \left\{-4\beta^{2}\lambda \right\}}{e^{6}\sqrt{\lambda}} \cdot n^{\frac{5}{8}} \delta_{n} \cdot \exp \left\{ -\frac{2\beta}{\lambda} n^{\frac{5}{4}} \delta_{n}^{2} \right\}.
\end{align}
Let us choose $\delta_{n}=\frac{1}{n^{5/8}}$, and then
\begin{align}
Q_{n}^{0}
&\geq \frac{1}{2} + \frac{(\pi\beta)^{\frac{3}{2}} \exp\{-4/\lambda\} \exp\left\{-4\beta^{2}\lambda \right\}}{e^{6}\sqrt{\lambda}} \cdot n^{\frac{5}{8}} \frac{1}{n^{5/8}} \cdot \exp \left\{ -\frac{2\beta}{\lambda} n^{\frac{5}{4}} \frac{1}{n^{5/4}} \right\} \\
&= \frac{1}{2} + \frac{(\pi\beta)^{\frac{3}{2}} \exp\{-4/\lambda\} \exp\left\{-4\beta^{2}\lambda \right\}\exp \left\{ -2\beta/\lambda \right\}}{e^{6}\sqrt{\lambda}} \\
&= \frac{1}{2} + \frac{(\pi\beta)^{\frac{3}{2}} \exp\{-(4+2\beta)/\lambda\} \exp\left\{-4\beta^{2}\lambda\right\}}{e^{6}\sqrt{\lambda}},
\end{align}
which is strictly larger then $\tfrac{1}{2}$, for any $\beta>0$ and $\lambda>0$.
Proposition \ref{PROP_4} is now proved.

\section*{Appendix F - Proof of Proposition \ref{PROP_6}}
\renewcommand{\theequation}{F.\arabic{equation}}
\setcounter{equation}{0}  

It follows from the union bound that 
\begin{align}
\pr\left\{N(\bX_{1};0) < 2n \right\}
&= \pr \left\{\bigcup_{i=1}^{2n} \{\bX_{1}(i) = 1\} \right\} \\
\label{ToSubs1}
&\leq \sum_{i=1}^{2n} \pr \left\{ \bX_{1}(i) = 1 \right\}.
\end{align} 

As before, let us denote $p_{n} = \tfrac{\lambda}{\sqrt{n}}$. Define the sequence $\{A_{n}\}_{n=1}^{\infty}$ by $A_{n}=\gamma n$, where $\gamma \in (0,1)$.
If an agent starts with a `0', then the probability to decide in favor of `1' is upper-bounded by 
\begin{align}
\label{B_ToExp0}
&\pr\left\{\text{Bin}\left(n-A_{n},p_{n}\right) \geq \text{Bin}\left(n+A_{n}-1,p_{n}\right) +1+1 \right\}  \\
\label{B_ToExp1}
&~~\leq \pr\left\{\text{Bin}\left(n-A_{n},p_{n}\right) \geq \text{Bin}\left(n+A_{n}-1,p_{n}\right) + \text{Bin}\left(1,p_{n}\right) \right\} \\
\label{ToCallA0}
&~~= \pr\left\{\text{Bin}\left(n-A_{n},p_{n}\right) \geq \text{Bin}\left(n+A_{n},p_{n}\right) \right\},
\end{align}
where the addition of the second 1 in \eqref{B_ToExp0} follows from the need to strictly break the tie in order to adopt `1' and \eqref{B_ToExp1} is due to the fact that $\text{Bin}\left(1,p_{n}\right) \leq 2$ with probability one. 

If an agent starts with a `1', then the probability to decide `1' is upper-bounded by 
\begin{align}
&\pr\left\{\text{Bin}\left(n-A_{n}-1,p_{n}\right) + 1 \geq \text{Bin}\left(n+A_{n},p_{n}\right) \right\} \nn \\
\label{ToCallA1}
&~~\leq \pr\left\{\text{Bin}\left(n-A_{n},p_{n}\right) + 1 \geq \text{Bin}\left(n+A_{n},p_{n}\right) \right\}.
\end{align}
Since \eqref{ToCallA1} cannot be smaller than \eqref{ToCallA0}, we continue with \eqref{ToCallA1}.
From now on, we prove that the probability in \eqref{ToCallA1}, to be denoted by $P_{n}$, converges to zero as $n \to \infty$.
Let
\begin{align}
X_{n} = \sum_{\ell=1}^{n-A_{n}} I_{\ell},~~~ 
Y_{n} = \sum_{k=1}^{n+A_{n}} J_{k},
\end{align}
where $I_{\ell} \sim \text{Ber}(p_{n})$, for all $\ell \in \{1,2,\ldots, n-A_{n}\}$, $J_{k} \sim \text{Ber}(p_{n})$, for all $k \in \{1,2,\ldots, n+A_{n}\}$, and all of these binary random variables are independent.
Now,
\begin{align}
P_{n} 
&= \pr \{X_{n}+1 \geq Y_{n}\} \\
&= \pr \left\{e^{\lambda(X_{n}-Y_{n}+1)} \geq 1 \right\} \\
\label{ToCallA2}
&\leq \mathbb{E} \left[e^{\lambda(X_{n}-Y_{n}+1)}\right],  
\end{align}
where \eqref{ToCallA2} is due to Markov's inequality. We get that
\begin{align}
\mathbb{E} \left[e^{\lambda(X_{n}-Y_{n}+1)}\right]
&= e^{\lambda} \cdot \mathbb{E}\left[\exp\left\{\lambda \left(\sum_{\ell=1}^{n-A_{n}} I_{\ell} - \sum_{k=1}^{n+A_{n}} J_{k}\right) \right\}\right] \\
&= e^{\lambda} \cdot \mathbb{E}\left[\prod_{\ell=1}^{n-A_{n}} e^{\lambda I_{\ell}} \cdot \prod_{k=1}^{n+A_{n}} e^{-\lambda J_{k}} \right] \\
\label{ToCallA3}
&= e^{\lambda} \cdot \prod_{\ell=1}^{n-A_{n}} \mathbb{E}\left[ e^{\lambda I_{\ell}} \right] \cdot \prod_{k=1}^{n+A_{n}} \mathbb{E}\left[ e^{-\lambda J_{k}} \right] \\
&= e^{\lambda} \cdot  \left[1+p_{n}(e^{\lambda}-1)\right]^{n-A_{n}} \cdot \left[1+p_{n}(e^{-\lambda}-1)\right]^{n+A_{n}}  \\
\label{ToCallA4}
&\leq e^{\lambda} \cdot  \left[\exp\{p_{n}(e^{\lambda}-1) \}\right]^{n-A_{n}} \cdot \left[\exp\{p_{n}(e^{-\lambda}-1)\}\right]^{n+A_{n}}  \\
\label{ToRef19}
&= \exp \left\{\lambda +p_{n}(e^{\lambda}-1)(n-A_{n}) +p_{n}(e^{-\lambda}-1)(n+A_{n}) \right\},
\end{align}
where \eqref{ToCallA3} is due to the independence of all binary random variables and \eqref{ToCallA4} follows from the inequality $1+x \leq e^{x}$. 
Since the bound in \eqref{ToRef19} is true for any $\lambda \geq 0$, it holds in particular for the choice
\begin{align}
e^{\lambda^{*}} = \sqrt{\frac{n+A_{n}}{n-A_{n}}}.
\end{align}
Substituting it back into \eqref{ToRef19} yields that
\begin{align}
P_{n} 
&\leq \exp \left\{\lambda^{*} +p_{n}(e^{\lambda^{*}}-1)(n-A_{n}) +p_{n}(e^{-\lambda^{*}}-1)(n+A_{n}) \right\} \\
&= \sqrt{\frac{n+A_{n}}{n-A_{n}}} \cdot \exp \left\{p_{n}\left(\sqrt{\frac{n+A_{n}}{n-A_{n}}}-1\right)(n-A_{n}) +p_{n}\left(\sqrt{\frac{n-A_{n}}{n+A_{n}}}-1\right)(n+A_{n}) \right\} \\
&= \sqrt{\frac{n+A_{n}}{n-A_{n}}} \cdot \exp \left\{p_{n}\left[\sqrt{(n+A_{n})(n-A_{n})}-n+A_{n}\right]  \right\} \nn \\
&~~~~~~~~~~~~~~~~~~~~~~~~~~~\times \exp \left\{p_{n}\left[\sqrt{(n-A_{n})(n+A_{n})}-n-A_{n}\right]  \right\} \\
&= \sqrt{\frac{n+A_{n}}{n-A_{n}}} \cdot \exp \left\{2p_{n}\left(\sqrt{n^{2}-A_{n}^{2}}-n\right) \right\}.
\end{align}
Consider the following
\begin{align}
\sqrt{n^{2}-A_{n}^{2}}-n
&= \sqrt{n^{2} \left(1-\frac{A_{n}^{2}}{n^{2}}\right)}-n \\
&= n \sqrt{1-\frac{A_{n}^{2}}{n^{2}}}-n \\
\label{ToCallA5}
&\leq n \left(1-\frac{A_{n}^{2}}{2n^{2}}\right)-n \\
&= -\frac{A_{n}^{2}}{2n},
\end{align}
where \eqref{ToCallA5} follows from the inequality $\sqrt{1-t} \leq 1-t/2$. Continuing from \eqref{ToSubs1}, we arrive at
\begin{align}
\pr\left\{N(\bX_{1};0) < 2n \right\}
&\leq 2n \sqrt{\frac{n+A_{n}}{n-A_{n}}} \cdot \exp \left\{-p_{n} \cdot \frac{A_{n}^{2}}{n} \right\},
\end{align}
and specifically, since we defined $A_{n} = \gamma n$, $\gamma \in (0,1)$, we arrive at
\begin{align}
\pr\left\{N(\bX_{1};0) < 2n \right\}
&\leq 2n \sqrt{\frac{n+\gamma n}{n-\gamma n}} \cdot \exp \left\{- \frac{\lambda}{\sqrt{n}} \cdot \frac{\gamma^{2} n^{2}}{n} \right\} \\
&= 2n \sqrt{\frac{1+\gamma}{1-\gamma}} \cdot \exp \left\{-\lambda \gamma^{2} \sqrt{n} \right\} \xrightarrow{n \to \infty} 0,
\end{align}
which completes the proof of Proposition \ref{PROP_6}.

\section*{Appendix G - Proof of Proposition \ref{PROP_7}}
\renewcommand{\theequation}{G.\arabic{equation}}
\setcounter{equation}{0}

Let us denote $N=N(\bX_{1};0)$.
For any $\mu \geq 0$, it follows from Markov's inequality that
\begin{align}
\pr \left\{N \geq n + B_{n} \right\}
&= \pr \left\{e^{\mu N} \geq e^{\mu(n + B_{n})} \right\} \\
\label{D_ToRef1}
&\leq \frac{\mathbb{E}\left[e^{\mu N}\right]}{e^{\mu(n + B_{n})}},
\end{align}
and thus, since \eqref{D_ToRef1} holds for every $\mu \geq 0$, it follows that
\begin{align}
\label{ToRef9}
\pr \left\{N \geq n + B_{n} \right\}
&\leq \inf_{\mu > 0} \frac{\mathbb{E}\left[e^{\mu N}\right]}{e^{\mu(n + B_{n})}}.
\end{align}
Note that
\begin{align}
N = \sum_{m=1}^{n+A_{n}} I_{m} + \sum_{m=1}^{n-A_{n}} J_{m},
\end{align}
where $I_{m} \sim \text{Ber}(P_{n,0})$, for all $m \in \{1,2,\ldots, n+A_{n}\}$, and $J_{m} \sim \text{Ber}(P_{n,1})$, for all $m \in \{1,2,\ldots, n-A_{n}\}$, and all of these binary random variables are independent.
We get that
\begin{align}
\mathbb{E}\left[e^{\mu N}\right]
&= \mathbb{E}\left[\exp\left\{\mu \left(\sum_{m=1}^{n+A_{n}} I_{m} + \sum_{m=1}^{n-A_{n}} J_{m}\right) \right\}\right] \\
&= \mathbb{E}\left[\prod_{m=1}^{n+A_{n}} e^{\mu I_{m}} \cdot \prod_{m=1}^{n-A_{n}} e^{\mu J_{m}} \right] \\
\label{D_ToRef2}
&= \prod_{m=1}^{n+A_{n}} \mathbb{E}\left[ e^{\mu I_{m}} \right] \cdot \prod_{m=1}^{n-A_{n}} \mathbb{E}\left[ e^{\mu J_{m}} \right] \\
&= \left(1-P_{n,0} + P_{n,0}e^{\mu}\right)^{n+A_{n}} \cdot \left(1-P_{n,1} + P_{n,1}e^{\mu}\right)^{n-A_{n}} \\
\label{D_ToRef3}
&= \left[1+P_{n,0}(e^{\mu}-1)\right]^{n+A_{n}} \cdot \left[1+P_{n,1}(e^{\mu}-1)\right]^{n-A_{n}} \\
\label{D_ToRef5}
&\leq \left[1+P_{n}(e^{\mu}-1)\right]^{n+A_{n}} \cdot \left[1+P_{n}(e^{\mu}-1)\right]^{n-A_{n}} \\ 
\label{D_ToRef6}
&= \left[1+P_{n}(e^{\mu}-1)\right]^{2n}.
\end{align}
where \eqref{D_ToRef2} is due to the independence of all binary random variables and \eqref{D_ToRef5} follows from the fact that the probability to update to `0' is upper-bounded by $P_{n}$.

Substituting \eqref{D_ToRef6} back into \eqref{ToRef9} yields that
\begin{align}
\pr \left\{N \geq n + B_{n} \right\}
&\leq \inf_{\mu > 0} \frac{\left[1+P_{n}(e^{\mu}-1)\right]^{2n}}{\exp\{\mu(n + B_{n})\}} \\
&= \inf_{\mu > 0} \exp \left\{2n \log \left[1+P_{n}(e^{\mu}-1)\right] - \mu(n + B_{n})\right\} \\
\label{ToRef10}
&= \exp \left\{\inf_{\mu > 0} \{2n \log \left[1+P_{n}(e^{\mu}-1)\right]-\mu(n + B_{n}) \} \right\}.
\end{align}
Upon defining 
\begin{align}
f(\mu) = 2n \log \left[1+P_{n}(e^{\mu}-1)\right]-\mu(n + B_{n}),
\end{align}
we find that the solution to $f'(\mu)=0$ is given by 
\begin{align}
\mu^{*}
= \log \left(\frac{\left(\frac{1}{2} + \frac{B_{n}}{2n}\right)\cdot(1-P_{n})}{\left(\frac{1}{2} - \frac{B_{n}}{2n}\right)\cdot P_{n}}\right).
\end{align}
Substituting it back into \eqref{ToRef10} provides that
\begin{align}
\pr \left\{N \geq n + B_{n} \right\} 
&\leq \exp \left\{2n \log \left[1+P_{n}(e^{\mu^{*}}-1)\right]-\mu^{*}(n + B_{n}) \right\} \\
&= \exp \left\{-2n \cdot \left[\left(\frac{1}{2} + \frac{B_{n}}{2n}\right)\log\frac{\frac{1}{2} + \frac{B_{n}}{2n}}{P_{n}} + \left(\frac{1}{2} - \frac{B_{n}}{2n}\right)\log\frac{\frac{1}{2} - \frac{B_{n}}{2n}}{1-P_{n}} \right] \right\} \\
&= \exp \left\{-2n \cdot D\left(\frac{1}{2} + \frac{B_{n}}{2n} \middle\| P_{n} \right) \right\}.
\end{align} 
which completes the proof of Proposition \ref{PROP_7}.

\section*{Appendix H - Proof of Proposition \ref{PROP_8}}
\renewcommand{\theequation}{H.\arabic{equation}}
\setcounter{equation}{0}

Assume that the numbers of zeros and ones are $n+\psi_{n}$ and $n-\psi_{n}$, respectively. 
If an agent starts with a `0', then the probability to decide in favor of `0' is upper-bounded by 
\begin{align}
&\pr\left\{\text{Bin}\left(n+\psi_{n}-1,p_{n}\right) + 1 \geq \text{Bin}\left(n-\psi_{n},p_{n}\right) \right\} \\
\label{ref40}
&~~~\leq \pr\left\{\text{Bin}\left(n+\psi_{n},p_{n}\right) + 1 \geq \text{Bin}\left(n-\psi_{n},p_{n}\right) \right\}.
\end{align}
If an agent starts with a `1', then the probability to decide in favor of `0' is upper-bounded by 
\begin{align}
&\pr\left\{\text{Bin}\left(n+\psi_{n},p_{n}\right) \geq \text{Bin}\left(n-\psi_{n}-1,p_{n}\right) + 1 + 1 \right\} \\
&~~~\leq \pr\left\{\text{Bin}\left(n+\psi_{n},p_{n}\right) \geq \text{Bin}\left(n-\psi_{n}-1,p_{n}\right) + \text{Bin}\left(1,p_{n}\right) \right\} \\
\label{ref41}
&~~~= \pr\left\{\text{Bin}\left(n+\psi_{n},p_{n}\right) \geq \text{Bin}\left(n-\psi_{n},p_{n}\right) \right\}.
\end{align}
Since \eqref{ref41} cannot be larger than \eqref{ref40}, we continue with \eqref{ref40}.
From now on, we upper-bound the probability in \eqref{ref40}, to be denoted by $P_{n}$.

Note that
\begin{align}
P_{n}
&= \pr\left\{\text{Bin}\left(n+\psi_{n},p_{n}\right) + 1 \geq \text{Bin}\left(n-\psi_{n},p_{n}\right) \right\} \\
&= \pr\left\{\text{Bin}\left(n-\psi_{n},p_{n}\right) + \text{Bin}\left(2\psi_{n},p_{n}\right) + 1 \geq \text{Bin}\left(n-\psi_{n},p_{n}\right) \right\} \\ 
&= \pr\left\{X + Z + 1 \geq Y \right\},
\end{align}
where $X,Y \sim \text{Bin}\left(n-\psi_{n},p_{n}\right)$ and $Z \sim \text{Bin}\left(2\psi_{n},p_{n}\right)$.
It follows from the law of total probability that 
\begin{align}
\pr\left\{X + Z + 1 \geq Y \right\}
\label{ref44}
&= \sum_{m=0}^{2\psi_{n}} \pr\left\{X + m + 1 \geq Y \right\} \pr\left\{Z = m \right\}.
\end{align}
As for the probability $\pr\left\{X + m + 1 \geq Y \right\}$, we have that 
\begin{align}
\pr\left\{X + m + 1 \geq Y \right\} 
\label{ref42}
= \pr\left\{X \geq Y \right\} + \pr\left\{X + 1 = Y \right\} + \ldots + \pr\left\{X + m + 1 = Y \right\}.
\end{align}
Substituting \eqref{ref1} back into \eqref{ref42} yields 
\begin{align}
&\pr\left\{X + m + 1 \geq Y \right\} \nn \\
&~~~~~~~~= \frac{1}{2} + \frac{1}{2} \cdot \pr\{X = Y\} + \pr\left\{X + 1 = Y \right\} + \ldots + \pr\left\{X + m + 1 = Y \right\} \\
\label{ref43}
&~~~~~~~~\leq \frac{1}{2} + \sum_{i=0}^{m+1} \pr\{X + i = Y \}.
\end{align}
As for the summands in \eqref{ref43}, we have that
\begin{align}
\pr\left\{X + i = Y \right\}
&= \sum_{\ell=0}^{N-i} \pr\{X=\ell\} \cdot \pr\{Y=\ell+i\} \\
\label{ToExp0}
&\leq \sqrt{\sum_{\ell=0}^{N-i} \left(\pr\{X=\ell\}\right)^{2}}
\sqrt{\sum_{\ell=0}^{N-i} \left(\pr\{Y=\ell+i\}\right)^{2}} \\
&= \sqrt{\sum_{\ell=0}^{N-i} \left(\pr\{X=\ell\}\right)^{2}}
\sqrt{\sum_{\ell=i}^{N} \left(\pr\{Y=\ell\}\right)^{2}} \\
&\leq \sqrt{\sum_{\ell=0}^{N} \left(\pr\{X=\ell\}\right)^{2}}
\sqrt{\sum_{\ell=0}^{N} \left(\pr\{Y=\ell\}\right)^{2}} \\
&=\sum_{\ell=0}^{N} \left(\pr\{X=\ell\}\right)^{2} \\
&=\sum_{\ell=0}^{N} \pr\{X=\ell\} \cdot \pr\{Y=\ell\}\\
\label{ref46}
&= \pr\{X=Y\},
\end{align} 
where \eqref{ToExp0} follows from the Cauchy-Schwarz inequality. Substituting \eqref{ref46} back into \eqref{ref43} yields that
\begin{align}
\pr\left\{X + m + 1 \geq Y \right\} 
&\leq \frac{1}{2} + \sum_{i=0}^{m+1} \pr\{X = Y\} \\
\label{ref47}
&= \frac{1}{2} + (m+2)\pr\{X = Y\}.
\end{align}
Upper-bounding \eqref{ref44} with \eqref{ref47} yields
\begin{align}
P_{n}
&= \sum_{m=0}^{2\psi_{n}} \pr\left\{X + m + 1 \geq Y \right\} \pr\left\{Z = m \right\} \\
&\leq \sum_{m=0}^{2\psi_{n}} \left[ \frac{1}{2} + (m+2)\pr\{X = Y\} \right] \pr\left\{Z = m \right\} \\
&= \frac{1}{2} + \sum_{m=0}^{2\psi_{n}} (m+2)\pr\{X = Y\} \pr\left\{Z = m \right\} \\
&= \frac{1}{2} + \pr\{X = Y\}(\mathbb{E}[Z]+2) \\
\label{ref45}
&= \frac{1}{2} + \pr\{X = Y\}(2\psi_{n}p_{n}+2).
\end{align}
The following result, which is proved in Appendix I, is analogues to Lemma \ref{Lemma_Binomial1} in Appendix B.
\begin{lemma} \label{Lemma_Binomial2}
	Let $X,Y \sim \text{Bin}\left(N,p_{n}\right)$, $N=n-\psi_{n}$, be two independent binomial random variables with $p_{n} = \tfrac{\lambda}{\sqrt{n}}$. 
	Define the sequence $\{\delta_{n}\}_{n=1}^{\infty}$ according to 
	\begin{align} \label{Delta_Def}
	n^{\delta_{n}} = \sqrt{\log(n^{\theta})},~~\theta > 5.
	\end{align} 
	Then, for all sufficiently large $n$,
	\begin{align}
	\pr\{X = Y\}
	\leq \frac{15}{\lambda n^{1/4-\delta_{n}}}.
	\end{align}
\end{lemma}
Continuing from \eqref{ref45}, we arrive at
\begin{align}
P_{n} 
&\leq \frac{1}{2} + \frac{30(\psi_{n}p_{n}+1)}{\lambda n^{1/4-\delta_{n}}} \\
&\leq \frac{1}{2} + \frac{60\psi_{n}p_{n}}{\lambda n^{1/4-\delta_{n}}},
\end{align}
which proves Proposition \ref{PROP_8}.

\section*{Appendix I - Proof of Lemma \ref{Lemma_Binomial2}}
\renewcommand{\theequation}{I.\arabic{equation}}
\setcounter{equation}{0}

Consider the following:
\begin{align}
&\pr\{X = Y\} \nn \\
&~~~~=\sum_{\ell=0}^{N} \left(\pr\{X=\ell\}\right)^{2}\\
&~~~~=\sum_{\ell=0}^{N} \left[\binom{N}{\ell} p_{n}^{\ell} (1-p_{n})^{N-\ell} \right]^{2} \\
&~~~~= \left[\binom{N}{0} p_{n}^{0} (1-p_{n})^{N} \right]^{2}
+\sum_{\ell=1}^{N-1} \left[\binom{N}{\ell} p_{n}^{\ell} (1-p_{n})^{N-\ell} \right]^{2} 
+ \left[\binom{N}{N} p_{n}^{N} (1-p_{n})^{0} \right]^{2}\\
\label{ref70}
&~~~~= (1-p_{n})^{2N} 
+\sum_{\ell=1}^{N-1} \left[\binom{N}{\ell} p_{n}^{\ell} (1-p_{n})^{N-\ell} \right]^{2} + p_{n}^{2N}. 
\end{align}

We continue by upper-bounding the PMF of the binomial random variable $X=\text{Bin}(N,p)$, which is given by
\begin{align}
\label{ref60}
P_{X}(k) = \binom{N}{k} p^{k} (1-p)^{N-k},~~~k \in [0:N].
\end{align}
In order to upper-bound the binomial coefficient in \eqref{ref60}, we use the Stirling's bounds in
\begin{align}
\sqrt{2 \pi n} \cdot n^{n} \cdot e^{-n} 
\leq n! 
\leq e \sqrt{n} \cdot n^{n} \cdot e^{-n},~~n \geq 1, 
\end{align} 
and get that
\begin{align}
\binom{N}{k}
&= \frac{N!}{k! \cdot (N-k)!} \\
\label{ref61}
&\leq \frac{e}{2\pi} \sqrt{\frac{N}{k(N-k)}}
\exp \left\{ -N \left[\frac{k}{N} \log \left(\frac{k}{N}\right) 
+ \left(1-\frac{k}{N}\right) \log \left(1-\frac{k}{N}\right) \right] \right\}.
\end{align}
Since $e < 2\pi$, substituting \eqref{ref61} back into \eqref{ref60} yields that for any $k=1,2,\ldots,n-1$
\begin{align}
P_{X}(k) 
\label{ref62}
&\leq \sqrt{\frac{N}{k(N-k)}}
\exp \left\{ -N D\left(\frac{k}{N} \middle\| p \right) \right\}, 
\end{align}
where $D(\alpha \| \beta)$, for $\alpha,\beta \in [0,1]$, is defined in \eqref{DEF_Bin_DIVERGENCE}.

As for the middle term in \eqref{ref70}, it follows from \eqref{ref62} that
\begin{align}
\sum_{\ell=1}^{N-1} \left[\binom{N}{\ell} p_{n}^{\ell} (1-p_{n})^{N-\ell} \right]^{2}
\label{ref71}
&\leq \sum_{\ell=1}^{N-1}  
\frac{N}{\ell(N-\ell)}
\exp \left\{ -2N D\left(\frac{\ell}{N} \middle\| p_{n} \right) \right\}.
\end{align}
In order to upper-bound \eqref{ref71}, 
let $\epsilon_{n} = \tfrac{1}{n^{3/4-\delta_{n}}}$, $n=1,2, \ldots$, where $\delta_{n} \to 0$ as $n \to \infty$, according to its definition in \eqref{Delta_Def}. Note that $\epsilon_{n}$ converges to zero faster than $p_{n}=\tfrac{\lambda}{\sqrt{n}}$ and define the set of numbers
\begin{align}
\calN_{n} = \{N(p_{n}-\epsilon_{n}),N(p_{n}-\epsilon_{n})+1,\ldots,Np_{n},\ldots,N(p_{n}+\epsilon_{n})\},
\end{align}
whose cardinality is given by
\begin{align}
\label{Cardinality3}
|\calN_{n}| = 2 N \epsilon_{n} +1.
\end{align}
Denote $\calM_{n} = \{1,2,\ldots,N-1\} \cap \calN_{n}^{\mbox{\tiny c}}$.
For any $\ell \in \calM_{n}$, it follows that
\begin{align}
D\left(\frac{\ell}{N} \middle\| p_{n} \right) 
&\geq D\left(p_{n}+\epsilon_{n} \middle\| p_{n} \right)\\
\label{ref75}
&= (p_{n}+\epsilon_{n}) \log \left(\frac{p_{n}+\epsilon_{n}}{p_{n}}\right)
+ (1-p_{n}-\epsilon_{n}) \log \left(\frac{1-p_{n}-\epsilon_{n}}{1-p_{n}}\right) \\
\label{ref73}
&= \left(\frac{1}{\sqrt{n}}+\frac{1}{n^{3/4-\delta_{n}}}\right)\log\left(1+\frac{1}{n^{1/4-\delta_{n}}}\right) \nn \\
&~~~~~~+ \left(1-\frac{1}{\sqrt{n}}-\frac{1}{n^{3/4-\delta_{n}}}\right)\log\left(1-\frac{1}{n^{3/4-\delta_{n}}-n^{1/4-\delta_{n}}}\right),
\end{align}
where in \eqref{ref73} we have substituted $p_{n}=\tfrac{1}{\sqrt{n}}$, since the actual value of $\lambda$ is immaterial for the asymptotic behavior of \eqref{ref75} as $n \to \infty$.
In order to lower-bound \eqref{ref73}, let us use the facts that $\log(1+t) \geq t-\tfrac{t^{2}}{2}$ for all $t\geq0$ and $\log(1-t)\geq -t-t^{2}$ for all $t\geq0$ sufficiently small. We find that \eqref{ref73} is lower-bounded by
\begin{align}
&\left(\frac{1}{\sqrt{n}}+\frac{1}{n^{3/4-\delta_{n}}}\right)\left(\frac{1}{n^{1/4-\delta_{n}}}-\frac{1}{2n^{1/2-2\delta_{n}}}\right) \nn \\
&~~~~~- \left(1-\frac{1}{\sqrt{n}}-\frac{1}{n^{3/4-\delta_{n}}}\right)\left(\frac{1}{n^{3/4-\delta_{n}}-n^{1/4-\delta_{n}}} + \frac{1}{(n^{3/4-\delta_{n}}-n^{1/4-\delta_{n}})^{2}}\right),
\end{align}  
which simplifies to 
\begin{align}
&\frac{n^{2\delta_{n}}-n^{3\delta_{n}-1/4}-2n^{2\delta_{n}-1/2}+2n^{3\delta_{n}-3/4}+n^{2\delta_{n}-1}+n^{3\delta_{n}-5/4}}{2n - 4\sqrt{n} + 2}  \\
&~~~\geq \frac{n^{2\delta_{n}}-n^{3\delta_{n}-1/4}-2n^{2\delta_{n}-1/2}}{2n - 4\sqrt{n} + 2} \\
&~~~\geq \frac{n^{2\delta_{n}}-\tfrac{1}{4}n^{2\delta_{n}}-\tfrac{1}{4}n^{2\delta_{n}}}{2n} \\
\label{ref72}
&~~~= \frac{n^{2\delta_{n}}}{4n} \\
&~~~\dfn \xi_{n}.
\end{align}
We now continue from \eqref{ref71} and arrive at
\begin{align}
&\sum_{\ell=1}^{N-1} 
\frac{N}{\ell(N-\ell)}
\exp \left\{ -2N D\left(\frac{\ell}{N} \middle\| p_{n} \right) \right\} \nn \\
\label{C_ToExp2}
&~~~\leq 
\sum_{\ell \in \calM_{n}} 
\frac{N}{\ell(N-\ell)}
\exp \left\{ -2N \xi_{n} \right\}
+
\sum_{\ell \in \calN_{n}} \frac{N}{\ell(N-\ell)} \\
&~~~\leq 
\sum_{\ell \in \calM_{n}} 
\frac{N}{(N-1)}
\exp \left\{ -2N \xi_{n} \right\}
+
\sum_{\ell \in \calN_{n}}  \frac{N}{N(p_{n}-\epsilon_{n})[N-N(p_{n}-\epsilon_{n})]} \\
\label{C_ToExp3}
&~~~\leq  
N \exp \left\{ -2N \xi_{n} \right\}
+ 
\frac{2N \epsilon_{n} +1}{N(1-p_{n}+\epsilon_{n})(p_{n}-\epsilon_{n})} \\
\label{ToRef5}
&~~~\leq  
N \exp \left\{ -2N \xi_{n} \right\}
+ 
\frac{2N \epsilon_{n} +N \epsilon_{n}}{N(1-\frac{1}{2})(p_{n}-\tfrac{1}{2}p_{n})} \\
\label{ToRef6}
&~~~\leq  
N \exp \left\{ -2N \xi_{n} \right\}
+ \frac{12\epsilon_{n}}{p_{n}}.
\end{align}
where \eqref{C_ToExp2} follows from the lower bound in \eqref{ref72} and the fact that $D(\alpha\|\beta) \geq 0$ in general. The passage to \eqref{C_ToExp3} is due to the fact that $|\calM_{n}| \leq N-1$ as well as \eqref{Cardinality3} and in \eqref{ToRef5}, we upper-bounded $1$ by $N\epsilon_{n}$, $p_{n}-\epsilon_{n}$ by $\tfrac{1}{2}$, and $\epsilon_{n}$ by $\tfrac{1}{2}p_{n}$.   

Upper-bounding \eqref{ref70} with \eqref{ToRef6} yields that
\begin{align}
\pr\{X = Y\}
\leq (1-p_{n})^{2N} 
+ N \exp \left\{ -2N \xi_{n} \right\}
+ \frac{12\epsilon_{n}}{p_{n}} + p_{n}^{2N}. 
\end{align}
Substituting $N=n-\psi_{n}$ and using the fact that $\lim_{n \to \infty} \frac{\psi_{n}}{n} = 0$ implies that for all sufficiently large $n$,
\begin{align}
&\pr\{X = Y\} \nn \\
&~~~~\leq \left(1-\frac{\lambda}{\sqrt{n}}\right)^{2(n-\psi_{n})}
+ (n-\psi_{n}) \exp \left\{ -2(n-\psi_{n})\xi_{n} \right\}
+ \frac{12}{\lambda n^{1/4-\delta_{n}}} + \left(\frac{\lambda}{\sqrt{n}}\right)^{2(n-\psi_{n})} \\
&~~~~\leq \left(1-\frac{\lambda}{\sqrt{n}}\right)^{2(n-\tfrac{n}{2})}
+ n \exp \left\{ -2(n-\tfrac{n}{2})\xi_{n} \right\}
+ \frac{12}{\lambda n^{1/4-\delta_{n}}} + \left(\frac{\lambda}{\sqrt{n}}\right)^{2(n-\tfrac{n}{2})} \\
&~~~~= \exp\left\{n \cdot \log \left(1-\frac{\lambda}{\sqrt{n}}\right) \right\} 
+ n \exp \left\{ -n\xi_{n} \right\}
+ \frac{12}{\lambda n^{1/4-\delta_{n}}} + \left(\frac{\lambda}{\sqrt{n}}\right)^{n}.
\end{align}
Substituting the expression of $\xi_{n}$ from \eqref{ref72} yields
\begin{align}
\label{ref74}
\pr\{X = Y\} 
&\leq 
\frac{12}{\lambda n^{1/4-\delta_{n}}}
+\exp\left\{n \cdot \log \left(1-\frac{\lambda}{\sqrt{n}}\right) \right\} + \left(\frac{\lambda}{\sqrt{n}}\right)^{n} + n \exp \left\{-\frac{n^{2\delta_{n}}}{4} \right\}.
\end{align}
For the specific choice $n^{\delta_{n}} = \sqrt{\log(n^{\theta})}$, the last term in \eqref{ref74} is given by
\begin{align}
n \exp \left\{-\frac{n^{2\delta_{n}}}{4} \right\}
&= n \exp \left\{-\frac{\log(n^{\theta})}{4} \right\} \\
&= \frac{1}{n^{\theta/4-1}}.
\end{align} 
Hence, if $\theta > 5$, there exists a sufficiently large $n$, such that the first term in \eqref{ref74} is larger than the other three terms, and hence,
\begin{align}
\pr\{X = Y\} 
\leq \frac{15}{\lambda n^{1/4-\delta_{n}}},
\end{align}
which completes the proof of Lemma \ref{Lemma_Binomial2}.

\section*{Appendix J - Proof of Proposition \ref{PROP_9}}
\renewcommand{\theequation}{J.\arabic{equation}}
\setcounter{equation}{0}  

\subsubsection*{Step 1: A Simplification for the Consensus Probability}
Due to symmetry, we only analyze the case $\mathsf{I}_{0} > \mathsf{I}_{1}$. 
It follows that 
\begin{align}
\pr\{\calC_{n}\} 
&= \pr\{N(\bX_{1};0) = 2n\}\\
&= \pr \left\{\bigcap_{i=1}^{2n} \{\bX_{1}(i) = 0\} \right\} \\
&= \prod_{i=1}^{2n} \pr \left\{ \bX_{1}(i) = 0 \right\} \\
\label{ToSubs2}
&= \prod_{i=1}^{2n} \left(1-\pr\left\{ \bX_{1}(i) = 1 \right\}\right).
\end{align}

\subsubsection*{Step 2: A Lower Bound on $\pr\left\{ \bX_{1}(i) = 1 \right\}$} 
If an agent starts with a `0', then the probability to decide in favor of `1' is lower-bounded by 
\begin{align}
&\pr\left\{\text{Bin}\left(n-C_{n},p_{n}\right) \geq \text{Bin}\left(n+C_{n}-1,p_{n}\right) +2 \right\} \nn \\
\label{ToCallE0}
&~~\geq \pr\left\{\text{Bin}\left(n-C_{n},p_{n}\right) \geq \text{Bin}\left(n+C_{n},p_{n}\right) + 2 \right\}.
\end{align} 
If an agent starts with a `1', then the probability to decide in favor of `1' is lower-bounded by 
\begin{align}
&\pr\left\{\text{Bin}\left(n-C_{n}-1,p_{n}\right) + 1 \geq \text{Bin}\left(n+C_{n},p_{n}\right) \right\} \nn \\
&~~\geq \pr\left\{\text{Bin}\left(n-C_{n}-1,p_{n}\right) + \text{Bin}\left(1,p_{n}\right) \geq \text{Bin}\left(n+C_{n},p_{n}\right) \right\}  \\
\label{ToCallE1}
&~~= \pr\left\{\text{Bin}\left(n-C_{n},p_{n}\right) \geq \text{Bin}\left(n+C_{n},p_{n}\right) \right\}.
\end{align}
Since \eqref{ToCallE0} cannot be larger than \eqref{ToCallE1}, we continue with \eqref{ToCallE0}.
From now on, we lower-bound the probability in \eqref{ToCallE0}, to be denoted by $Q_{n}$.
The probability in \eqref{ToCallE0} can be written explicitly as 
\begin{align}
\label{ToCont10}
Q_{n} = \sum_{\ell=0}^{n-C_{n}} \sum_{k=0}^{n+C_{n}}
\binom{n-C_{n}}{\ell} p_{n}^{\ell} (1-p_{n})^{n-C_{n}-\ell}
\binom{n+C_{n}}{k} p_{n}^{k} (1-p_{n})^{n+C_{n}-k} \IND \{\ell \geq k+2\}.
\end{align} 
In order to continue, recall from \eqref{ToRef4} that
\begin{align}
P_{X}(k) 
\label{ToRef7}
&\geq \frac{\sqrt{2\pi}}{e^{2}} \sqrt{\frac{1}{k}}
\exp \left\{ -n D\left(\frac{k}{n} \middle\| p \right) \right\} ,
\end{align}
where $D(\alpha \| \beta)$, for $\alpha,\beta \in [0,1]$, is defined in \eqref{DEF_Bin_DIVERGENCE}.  
Substituting twice this lower bound into \eqref{ToCont10}, we arrive at  
\begin{align}
\label{ToCont1}
Q_{n} 
&\geq 
\frac{2\pi}{e^{4}}
\sum_{\ell=0}^{n-C_{n}} \sum_{k=0}^{\ell-2}
\sqrt{\frac{1}{\ell}}
\exp \left\{ -(n-C_{n}) D\left(\frac{\ell}{n-C_{n}} \middle\| p_{n} \right) \right\} \nn \\
&~~~~~~~~~~~~~~~~~~~~~\times 
\sqrt{\frac{1}{k}}
\exp \left\{ -(n+C_{n}) D\left(\frac{k}{n+C_{n}} \middle\| p_{n} \right) \right\} \\
\label{ToExp7}
&\geq 
\frac{8\pi}{e^{4}}
\sum_{\ell=(n-C_{n})p_{n}}^{(n+C_{n})p_{n}}
\sum_{k=0}^{\ell-2} \frac{1}{\sqrt{\ell k}}
\exp \left\{ -(n-C_{n}) D\left(\frac{\ell}{n-C_{n}} \middle\| p_{n} \right) \right\} \nn \\
&~~~~~~~~~~~~~~~~~~~~~~~~~~~~~~~~~~~~~~~~~~\times 
\exp \left\{ -(n+C_{n}) D\left(\frac{k}{n+C_{n}} \middle\| p_{n} \right) \right\} \\
\label{ToExp8}
&\geq 
\frac{8\pi}{e^{4}}
\sum_{\ell=(n-C_{n})p_{n}}^{(n+C_{n})p_{n}}
\sum_{k=\ell - C_{n}p_{n}}^{\ell-2} 
\frac{1}{\sqrt{\ell(\ell-2)}}
\exp \left\{ -(n-C_{n}) D\left(\frac{\ell}{n-C_{n}} \middle\| p_{n} \right) \right\} \nn \\
&~~~~~~~~~~~~~~~~~~~~~~~~~~~~~~~~~~~~~~~~~~\times 
\exp \left\{ -(n+C_{n}) D\left(\frac{k}{n+C_{n}} \middle\| p_{n} \right) \right\} \\
\label{ToExp9}
&\geq 
\frac{8\pi}{e^{4}(n+C_{n})p_{n}}
\sum_{\ell=(n-C_{n})p_{n}}^{(n+C_{n})p_{n}}
\sum_{j=2}^{C_{n}p_{n}}
\exp \left\{ -(n-C_{n}) D\left(\frac{\ell}{n-C_{n}} \middle\| p_{n} \right) \right\} \nn \\
&~~~~~~~~~~~~~~~~~~~~~~~~~~~~~~~~~~~~~~~~~~\times 
\exp \left\{ -(n+C_{n}) D\left(\frac{\ell-j}{n+C_{n}} \middle\| p_{n} \right) \right\} \\
\label{ToCont3}
&=
\frac{8\pi}{e^{4}(n+C_{n})p_{n}}
\sum_{m=0}^{2C_{n}} \sum_{j=2}^{C_{n}p_{n}}
\exp \left\{ -(n-C_{n}) D\left(\frac{(n-C_{n}+m)p_{n}}{n-C_{n}} \middle\| p_{n} \right) \right\} \nn \\
&~~~~~~~~~~~~~~~~~~~~~~~~~~~~~~~~~~~~~\times 
\exp \left\{ -(n+C_{n}) D\left(\frac{(n-C_{n}+m)p_{n}-j}{n+C_{n}} \middle\| p_{n} \right) \right\},
\end{align}
where \eqref{ToExp7} follows from the condition $\lim_{n \to \infty} C_{n}/n = 0$, which implies that for all large enough $n$, both $(n-C_{n})p_{n} \geq 0$ and $(n+C_{n})p_{n} \leq n-C_{n}$ hold.
The inequality in \eqref{ToExp8} follows from the fact that $\ell \geq C_{n}p_{n}$, for all sufficiently large $n$.
In \eqref{ToExp9} we changed the summation index from $k$ to $j$ according to $k=\ell-j$, with $j \in \{2,3,\ldots,C_{n}p_{n}\}$, and in \eqref{ToCont3} we changed the summation index from $\ell$ to $m$ according to $\ell=(n-C_{n}+m)p_{n}$, with $m \in \{0,1,\ldots,2C_{n}\}$. 
In order to upper-bound the divergence terms in \eqref{ToCont3}, we use the reverse Pinsker inequality in \eqref{Reverse_PINSKER}. 
Then, after some algebraic work, we arrive at
\begin{align}
Q_{n}
&\geq
\frac{8\pi}{e^{4}(n+C_{n})p_{n}}
\sum_{m=0}^{2C_{n}} \sum_{j=2}^{C_{n}p_{n}}
\exp \left\{ -(n-C_{n}) \cdot \frac{2}{p_{n}}  \frac{p_{n}^{2} m^{2}}{(n-C_{n})^{2}} \right\} \nn \\
&~~~~~~~~~~~~~~~~~~~~~~~~~~~~~~~~~~~~~~~~\times 
\exp \left\{ -(n+C_{n}) \cdot \frac{2}{p_{n}} \frac{[p_{n}(2C_{n}-m)+j]^{2}}{(n+C_{n})^{2}} \right\} \\
&=
\frac{8\pi}{e^{4}(n+C_{n})p_{n}}
\sum_{m=0}^{2C_{n}} \sum_{j=2}^{C_{n}p_{n}}
\exp \left\{ - 2 \frac{p_{n} m^{2}}{n-C_{n}} \right\} \cdot
\exp \left\{ - \frac{2}{p_{n}} \frac{[p_{n}(2C_{n}-m)+j]^{2}}{n+C_{n}} \right\} \\
\label{ToExp10}
&\geq
\frac{8\pi}{e^{4}(n+C_{n})p_{n}}
\sum_{m=0}^{2C_{n}} \sum_{j=2}^{C_{n}p_{n}}
\exp \left\{ -2  \frac{p_{n}m^{2}}{n-C_{n}} \right\} \cdot
\exp \left\{ - \frac{2}{p_{n}} \frac{[p_{n}(2C_{n}-m)+2p_{n}C_{n}]^{2}}{n-C_{n}} \right\} \\
&\geq
\frac{4\pi C_{n}p_{n}}{e^{4}(n+C_{n})p_{n}}
\sum_{m=0}^{2C_{n}}
\exp \left\{ -2  \frac{p_{n}m^{2}}{n-C_{n}} \right\} \cdot
\exp \left\{-2 \frac{p_{n}(4C_{n}-m)^{2}}{n-C_{n}} \right\} \\
&=
\frac{4\pi C_{n}}{e^{4}(n+C_{n})}
\sum_{m=0}^{2C_{n}}
\exp \left\{ - 2p_{n} \cdot  \frac{m^{2}+(4C_{n}-m)^{2}}{n-C_{n}} \right\} \\
\label{ToRef20}
&=
\frac{4\pi C_{n}}{e^{4}(n+C_{n})}
\sum_{m=0}^{2C_{n}}
\exp \left\{ - 4p_{n} \cdot  \frac{(2C_{n}-m)^{2} + 4C_{n}^{2}}{n-C_{n}} \right\},
\end{align}
where \eqref{ToExp10} is true since $n-C_{n} \leq n+C_{n}$, and due to the fact that $j$ is obviously upper-bounded by $2p_{n}C_{n}$. 
Now, the exponent in \eqref{ToRef20} is maximized at $m=0$, and thus
\begin{align}
Q_{n} 
&\geq \frac{4\pi C_{n}}{e^{4}(n+C_{n})}
\sum_{m=0}^{2C_{n}}
\exp \left\{ - 32p_{n} \cdot  \frac{C_{n}^{2}}{n-C_{n}} \right\} \\
&\geq \frac{8\pi C_{n}^{2}}{e^{4}(n+C_{n})}
\exp \left\{ - 32p_{n} \cdot  \frac{C_{n}^{2}}{n-C_{n}} \right\}.
\end{align}

\subsubsection*{Step 3: Wrapping Up}
Continuing from \eqref{ToSubs2}, we finally arrive at
\begin{align}
\pr\{\calC_{n}\} 
&\leq \prod_{i=1}^{2n} \left(1-\frac{8\pi C_{n}^{2}}{e^{4}(n+C_{n})}
\exp \left\{ - 32p_{n} \cdot  \frac{C_{n}^{2}}{n-C_{n}} \right\}\right) \\
&= \left(1 - \frac{8\pi C_{n}^{2}}{e^{4}(n+C_{n})}
\exp \left\{ - 32p_{n} \cdot  \frac{C_{n}^{2}}{n-C_{n}} \right\} \right)^{2n} \\
&= \exp\left\{ 2n \cdot \log\left(1 - \frac{8\pi C_{n}^{2}}{e^{4}(n+C_{n})}
\exp \left\{ - 32p_{n} \cdot  \frac{C_{n}^{2}}{n-C_{n}} \right\} \right) \right\} \\
\label{ToExp6}
&\leq \exp\left\{ - \frac{16\pi n C_{n}^{2}}{e^{4}(n+C_{n})}
\exp \left\{ - 32p_{n} \cdot  \frac{C_{n}^{2}}{n-C_{n}} \right\} \right\} \\
&\leq \exp\left\{ -\frac{n C_{n}^{2}}{2(n+C_{n})}
\exp \left\{ - 32p_{n} \cdot  \frac{C_{n}^{2}}{n-C_{n}} \right\} \right\}, 
\end{align}
where \eqref{ToExp6} is due to $\log(1-y) \leq -y$.
This completes the proof of Proposition \ref{PROP_9}.

\section*{Acknowledgment}

The author is grateful to Yonatan Shadmi (Technion) for reading the manuscript and providing some valuable comments, and especially for supplying the motivation to prove Theorem \ref{Main_THEOREM2}.


\begin{thebibliography}{AA}

\bibitem{Benjamini2016}
I.~Benjamini, S.-O.~Chan, R.~O'Donnell, O.~Tamuz, and L.-Y.~Tan, ``Convergence, unanimity and disagreement in majority dynamics on unimodular graphs and random graphs,'' {\it Stochastic Processes and their Applications}, vol. 126, no. 9, pp. 2719--2733, Sep.\ 2016.

\bibitem{Fountoulakis2020}
N.~Fountoulakis, M.~Kang, and T.~Makai, ``Resolution of a conjecture on majority dynamics: Rapid stabilization in dense random graphs,'' {\it Random Structures \& Algorithms}, vol. 57, no. 4, pp. 1134--1156, Oct.\ 2020. 


\bibitem{gray2006consensus} 
J. Gray and L. Lamport, ``Consensus on transaction commit,'' {\it ACM Transactions on
	Database Systems (TODS),} vol. 31, no. 1, pp. 133--160, 2006.

\bibitem{antoniadis2018state} 
K. Antoniadis, R. Guerraoui, D. Malkhi, and D.-A. Seredinschi, ``State machine replication
is more expensive than consensus,'' Tech. Rep., 2018.

\bibitem{attiya1998atomic} 
H. Attiya and O. Rachman, ``Atomic snapshots in $o(n log n)$ operations,'' {\it SIAM Journal
	on Computing,} vol. 27, no. 2, pp. 319--340, 1998.

\bibitem{lamport2019time} 
L. Lamport, ``Time, clocks, and the ordering of events in a distributed system,'' in {\it Concurrency: the Works of Leslie Lamport,} 2019, pp. 179--196.

\bibitem{mustafa2001majority} 
N. H. Mustafa and A. Peke{\v{c}}, ``Majority consensus and the local majority rule,'' in {\it International Colloquium on Automata, Languages, and Programming.} Springer, 2001, pp.	530--542.

\bibitem{moreira2004efficient} 
A. A. Moreira, A. Mathur, D. Diermeier, and L. A. Amaral, ``Efficient system-wide coordination in noisy environments,'' {\it Proceedings of the National Academy of Sciences,} vol. 101, no. 33, pp. 12085--12090, 2004.

\bibitem{gogolev2015distributed} A. Gogolev, N. Marchenko, L. Marcenaro, and C. Bettstetter, ``Distributed binary consensus in networks with disturbances,'' {\it ACM Transactions on Autonomous and Adaptive Systems (TAAS),} vol. 10, no. 3, pp. 1--17, 2015.

\bibitem{thomas1979majority} 
R. H. Thomas, ``A majority consensus approach to concurrency control for multiple copy databases,'' {\it ACM Transactions on Database Systems (TODS),} vol. 4, no. 2, pp. 180--209, 1979.

\bibitem{breitwieser1982distributed} 
H. Breitwieser and M. Leszak, ``A distributed transaction processing protocol based on majority consensus,'' in {\it Proceedings of the first ACM SIGACT-SIGOPS symposium on Principles of distributed computing,} 1982, pp. 224--237.

\bibitem{kanrar2016new} 
S. Kanrar, S. Chattopadhyay, and N. Chaki, ``A new hybrid mutual exclusion algorithm in the absence of majority consensus,'' in {\it Advanced Computing and Systems for Security.} Springer, 2016, pp. 201--214.

\bibitem{mostofi2007binary} 
Y. Mostofi, ``Binary consensus with gaussian communication noise: A probabilistic approach,'' in {\it 2007 46th IEEE Conference on Decision and Control.} IEEE, 2007, pp. 2528--2533.

\bibitem{perron2009using} 
E. Perron, D. Vasudevan, and M. Vojnovic, ``Using three states for binary consensus on complete graphs,'' in {\it IEEE INFOCOM 2009.} IEEE, 2009, pp. 2527--2535.

\bibitem{cruise2014probabilistic} 
J. Cruise and A. Ganesh, ``Probabilistic consensus via polling and majority rules,'' Queueing Systems, vol. 78, no. 2, pp. 99--120, 2014.

\bibitem{PAPER1}
M.~A.~Abdullah and M.~Draief, ``Global majority consensus by local majority polling on graphs of a given degree sequence,'' {\it Discret. Appl. Math.,} vol. 180, no. 10, pp. 1-10, Jan. 2015.

\bibitem{PAPER2}
B.~G\"artner and A.~N.~Zehmakan, ``Majority model on random regular graphs,'' In Proceedings of the {\it Latin American Symposium on Theoretical Informatics}, Buenos Aires, Argentina, April 16-19, 2018; pp. 572--583.

\bibitem{PAPER3}
L.~Tran and V.~Vu, ``Reaching a consensus on random networks: The power of few,'' {\it Approximation, Randomization, and Combinatorial Optimization. Algorithms and Techniques (APPROX/RANDOM 2020)}, 
Leibniz International Proceedings in Informatics, vol. 176, pp. 20:1-20:15, 2020. 

\bibitem{PAPER4}
A.~N.~Zehmakan, ``Opinion forming in Erd\H{o}s--R\'enyi random graph and expanders,'' {\it The 29th International Symposium on Algorithms and Computation (ISAAC2018)}, 
Leibniz International Proceedings in Informatics, vol. 168, pp. 4:1-4:13, 2018

\bibitem{TLS}
R.~Tamir, A.~Livshits, and Y.~Shadmi, ``Simple majority consensus in networks with unreliable communication,'' {\it Entropy} 24(3), 333. Published on February 25, 2022. {\tt https://doi.org/10.3390/e24030333}.

	
	\bibitem{DURRETT}
	R.~Durrett, {\it Probability: Theory and Examples}, Cambridge University Press, Fifth edition, 2019.
	
	\bibitem{SASON}
	I.~Sason and S.~Verd\'u, ``f-divergence inequalities,'' {\it IEEE Trans. on Information Theory}, vol. 62, no. 11, pp. 5973--6006, November 2016.
	
	\bibitem{C1967}
	I.~Csisz\'ar, ``Information-type measures of difference of probability distributions and indirect observations,'' {\it Studia Scientiarum Mathematicarum Hungarica,} vol.\ 2, pp.\ 299--318, Jan.\ 1967.
	
	\bibitem{Kullback}
	S.~Kullback, ``A lower bound for discrimination information in terms of variation,'' {\it IEEE Trans. on Information Theory}, vol. 13, no. 1, pp. 126--127, Jan.\ 1967.
	
\end{thebibliography}
\end{document}